\documentclass[usenatbib,useAMS]{mn2e}

\usepackage{amsfonts}
\usepackage{amsmath}
\usepackage{wasysym}
\usepackage{graphicx}
\usepackage{longtable}
\usepackage{lscape}
 
\begin{document}

\label{firstpage}

\title[catalogue of early-type galaxies]{An extensive catalogue of early-type galaxies in the nearby Universe}

\author[Dabringhausen \& Fellhauer]{
J. Dabringhausen$^{1}$ \thanks{E-mail: joerg@astro-udec.cl},
M. Fellhauer$^{1}$\\
$^{1}$ Departamento de Astronom\'{i}a, Universidad de Concepcion, Casilla 160-C, Concepcion, Chile}

\pagerange{\pageref{firstpage}--\pageref{lastpage}} \pubyear{2015}

\maketitle

\begin{abstract}
We present a catalogue of 1715 early-type galaxies from the literature, spanning the luminosity range from faint dwarf spheroidal galaxies to giant elliptical galaxies. The aim of this catalogue is to be one of the most comprehensive and publicly available collections of data on early-type galaxies. The emphasis in this catalogue lies on dwarf elliptical galaxies, for which some samples with detailed data have been published recently. For almost all of the early-type galaxies included in it, this catalogue contains data on their locations, distances, redshifts, half-light radii, the masses of their stellar populations and apparent magnitudes in various passbands. Data on metallicity and various colours are available for a majority of the galaxies presented here. The data on magnitudes, colours, metallicities and masses of the stellar populations is supplemented with entries that are based on fits to data from simple stellar population models and existing data from observations. Also, some simple transformations have been applied to the data on magnitudes, colours and metallicities in this catalog, in order to increase the homogeneity of this data. Estimates on the S\'{e}rsic profiles, internal velocity dispersions, maximum rotational velocities, dynamical masses and ages are listed for several hundreds of the galaxies in this catalogue. Finally, each quantity listed in this catalogue is accompanied with information on its source, so that users of this catalogue can easily exclude data that they do not consider as reliable enough for their purposes.
\end{abstract}


\section[Introduction]{Introduction}
\label{sec:introduction}

Early-type galaxies (ETGs) are galaxies in which recent star formation is negligible and in which random motion of the stars dominates over ordered motion. Apart from these two defining properties, the properties of ETGs are diverse. For instance, the ETGs span a range over 9 magnitudes in luminosity, and there are also noticeable differences in their colours, ages, metallicities, radii and internal velocity dispersions.

Investigating how these parameters are correlated with each other can provide valuable insights on the nature of the ETGs. The key ingredient for this kind of work are catalogs of galaxies that are large, but at the same time contain accurate information on the galaxies. In practise, a compromise needs to be found, so that both goals can be reached.

Since the pioneering works in that field (e.g. \citealt{Faber1976,Kormendy1985,Djorgovski1987,Bender1992}), the available information on ETGs has grown enormously. At the bright end of the luminosity function of ETGs, the ${\rm ATLAS}^{\rm 3D}$ project \citep{Cappellari2011} has provided a wealth of data for a sample of 260 ETGs that are within a distance of 42 Mpc of the Milky Way. The faint end of the luminosity function is covered with the detailed catalogue by \citet{McConnachie2012}, who collected data on all known galaxies within a distance of 3 Mpc of the Milky Way. ETGs in the intermediate luminosity range have been studied in great detail by, e.g., \citet{Geha2003}, \citet{Chilingarian2008}, \citet{Kourkchi2012b} and \citet{Toloba2014}.

For the catalogue presented here, we have searched the available literature on ETGs in order to collect as much and as detailed information as possible into a single catalogue that spans the whole luminosity range of ETGs. An emphasis was placed on the inclusion of dwarf elliptical galaxies (dEs), i.e. on galaxies filling the luminosity gap in between the sample of very faint ETGs taken from the catalogue on galaxies in the Local Group by \citet{McConnachie2012} and the sample of bright ETGs provided by the ATLAS$^{\rm 3D}$ project.

\section[Overview]{Overview}
\label{sec:overview}

Our census of the literature on ETGs has resulted in a sample of 1715 galaxies. For almost all of them, we provide data on locations, distances, redshifts, half-light radii, the masses of their stellar populations and apparent magnitudes in various passbands. Various colours are available for about 1000 ETGs in this catalogue. Also listed in this catalogue are S\'{e}rsic indices for almost all ETGs, internal velocity dispersions for 725 ETGs, maximum rotational velocities for 446 ETGs, dynamical masses for 723 ETGs, estimates for the characteristic metallicity for 1242 ETGs and estimates for the characteristic age for 467 ETGs.

The catalogue presented here contains two types of data. The first type of data are values that are obtained from dedicated observations of individual galaxies. Examples for this type of data are directly observed apparent magnitudes, colours derived from such magnitudes, or internal velocity dispersions derived from the spectra of the galaxies. The second type of data is values that are assigned to a galaxy based on the statistical properties of a group of similar galaxies for which data of the first type is available, or simple fits to Simple Stellar Population models (SSP-models). An example for this second type of data are luminosities in the $X$-band that are calculated from luminosities in the $Y$-band, using a fit that quantifies how $X$-band luminosities correlate with $Y$-band luminosities. Data of this second type is calculated for this catalogue if data of the first type is not available for the desired quantity.

Our catalogue merges data from many different teams, who obtained their raw data under different conditions with different instruments, and perhaps also reduced the raw data in different ways. As a consequence, the data in this catalogue is inherently inhomogeneous, even though we apply some basic transformations to the data where this seems possible and prudent, as specified in Sections~(\ref{sec:magnitudes}) to~(\ref{sec:properties}). We also inherit inhomogeneities from other catalogs, if data from them are included as subsamples of this catalogue. This concerns in particular the catalogue of galaxies in the Local Group by \citet{McConnachie2012}, which is for the catalogue presented here the primary source for the faintest ETGs. Most of these galaxies have been discovered only recently, are very challenging to observe, and the original data on them is distributed over many original papers. Nevertheless, the catalogue by \citet{McConnachie2012} still represents the state of the art on the information that is available for the galaxies at the faint end of the ETG luminosity function, even though the data provided for them are in many cases only proxies to the quantities they represent.

Being aware of the fact that the data presented in this catalogue cannot be fully homogeneous, we provide detailed information on the sources for the data. Thus, all data on parameters given for a galaxy are accompanied with data on the source for those parameters. This information is coded in numbers. In this context, numbers below 100 indicate that value for the given parameter for the given galaxy basically comes from the literature or a publicly accessible data base, even though this data might have been homogenised for the inclusion in this catalog, as specified in Sections~(\ref{sec:magnitudes}) to~(\ref{sec:properties}). For parameters that have been calculated based on the statistical properties of a sample of galaxies, or values that have been calculated for this catalogue based on SSP-models, the numbers encoding the provenience of the data are always larger than 100. This provides an easy-to-implement criterion for discarding such estimates.

In our catalogue, we prefer previously published data over estimates done for this catalogue. However, sometimes more than one value for some parameter for a given galaxy has been published before. In this case, the data from the newest source is usually preferred over data from older sources. This is motivated with the assumption that newer data should be superior over older data. Also, data coming from larger samples are preferred over data coming from smaller samples. This is done in an effort to have large subsamples in the catalogue that are at least internally homogeneous. Finally, data from subsamples with more comprehensive information are usually preferred over data from samples with less comprehensive information. For instance, data from a subsample on half-light radii and internal velocity dispersions would be preferred over a sample that only contains data on the half-light radii on the same galaxies, even if the latter is larger or more recent. These guidelines are supposed to make the catalogue presented here as up-to-date and as homogeneous as possible. Note, however, that these guidelines cannot be applied in a strict manner, since they are sometimes in conflict with each other.

The data is presented in 10 tables, in which the galaxies are ordered by ascending right ascension $\alpha(2000.0)$. The first table lists different names of the galaxies. The second table lists positions, distances and redshifts of the galaxies. The third and the fourth table list the apparent magnitudes of the galaxies in various passbands. The fifth and the sixth table list the luminosity of the galaxies in Solar units. The seventh and the eighth table list different colours of the galaxies. The ninth table lists half-light radii, S\'{e}rsic indices,  internal velocity dispersions, maximum rotation velocities and dynamical masses of the galaxies. The tenth table lists masses of the stellar populations, SSP-equivalent metallicities and SSP-equivalent ages of the galaxies. The tables are described in detail in the following sections.

These tables can be downloaded from the webpage of the theory group at the department of astronomy at the Universidad de Concepci\'{o}n (http://www.astro-udec.cl/mf/catalogue/).

\section[Names]{Names}
\label{sec:names}

\subsection{General Procedures}
\label{sec:names1}

The data on the galaxies in our catalogue come from many different sources, and our catalogue also includes many faint galaxies that have only been discovered very recently. For these reasons, there is no universal naming scheme for all of them. We therefore list in Tab.~(\ref{tab11}) different names by which the galaxies in our catalogue are known. For each galaxy in our catalog, there is at least one name provided in Tab.~(\ref{tab11}) by which it can be found either in the NASA/IPAC Extragalactic Database\footnote{http://ned.ipac.caltech.edu/} (NED), or the HyperLeda database\footnote{http://leda.univ-lyon1.fr/} \citep{Makarov2014}. The galaxies are ordered in Tab.~(\ref{tab11}) by ascending right ascension $\alpha(2000.0)$.

For quick and easy navigation between the different tables of this catalog, we provide each galaxy with a running number that identifies it. These numbers are repeated in the first column in each of the ten tables of this catalogue. However, these numbers are not intended for use outside this catalog, as they might change in future updates of this catalogue.

\subsection{Detailed description of the data}
\label{sec:names2}

In the following, we describe the entries in Tab.~(\ref{tab11}) column by column. Each column except for the last corresponds to a catalogue in which a given galaxy has been listed previously. The last column lists alternative names that do not appear in any of the other columns. The entries that table are set to 0, if a galaxy is not listed in the catalogue corresponding to a given column.

{\bf Column~1} lists the running number that identifies the galaxies within this catalogue. The galaxies are ordered by ascending right ascension $\alpha(2000.0)$.

{\bf Column~2} lists the identification number of the galaxy in the New General Catalog, NGC \citep{Dreyer1888}.

{\bf Column~3} lists the identification number of the galaxy in the Index Catalog, IC \citep{Dreyer1895}.

{\bf Column~4} lists the identification number of the galaxy in the Uppsala General Catalogue of Galaxies, UGC \citep{Cotton1999}.

{\bf Column~5} lists the identification number of the galaxy in the Arecibo General Catalog, AGC \citep{Scodeggio1998a}.

{\bf Column~6} lists the identification number of the galaxy in the Principal Galaxy Catalog, PGC. This is the primary identification used for the objects listed in HyperLeda \citep{Makarov2014}.

{\bf Column~7} lists the identification code of the galaxy in the Catalogue of Galaxies and of Clusters of Galaxies, CGCG \citep{Zwicky1968}.

{\bf Column~8} lists the identification number of the galaxy in the Virgo Cluster Catalog, VCC \citep{Binggeli1985}.

{\bf Column~9} lists the identification number of the galaxy in the Godwin+Metcalf+Peach catalogue of galaxies in clusters, GMP \citep{Godwin1983}.

{\bf Column~10} lists the identification number of the galaxy in the Fornax Cluster Catalog, FCC \citep{Ferguson1989}.

{\bf Column~11} lists the identification number of the galaxy in the catalogue of Virgo cluster galaxies by \citet{Lieder2012}, [LHH2012].

{\bf Column~12} lists the identification code of the galaxy in the catalogue of Centaurus cluster galaxies by \citet{Misgeld2009}, [MHM2009].

{\bf Column~13} lists the identification number of the galaxy in the Hydra Cluster Catalog, HCC \citep{Misgeld2008}.

{\bf Column~14} lists the identification number of the galaxy in the catalogue of galaxies in the galaxy cluster Abell~496 \citep{Chilingarian2008}.

{\bf Column~15} lists other names or designations for the galaxies, namely entries in the Messier catalogue and names given to dwarf galaxies in the Local Group. The dwarf galaxies in the Local Group are actually often only known by those names.

\begin{landscape}
\begin{table}
\caption{\label{tab11} Names of the galaxies in this catalogue. A portion of the Table is shown here for guidance regarding its contents and form. A detailed description of the contents of this table is given in Section~(\ref{sec:names2}). A detailed description of the data in the table is given in section~(\ref{sec:names2}) as well. The table in its entirety can be downloaded at http://www.astro-udec.cl/mf/catalogue/.}
\centering
\vspace{2mm}
\begin{tabular}{rrrrrrrrrrrrrrr}
\hline
&&&&&&&&&&&&&& \\[-10pt]
id     &       NGC   &   IC    &  UGC       &    AGC     &    PGC     &   CGCG     &   VCC    & GMP     &    FCC  & [LHH2012] & [MHM2009] & HCC & Abell~496 & other \\
\hline
$ 501$ & $   0 $ & $  560$ & $ 5223 $ & $       0$ & $   27998$ & $   7-030$ & $     0$ & $    0$ & $    0$ & $   0$ & $ 0 $ & $   0$ & $   0$ & $     0$\\
$ 502$ & $3032 $ & $    0$ & $ 5292 $ & $       0$ & $   28424$ & $ 152-077$ & $     0$ & $    0$ & $    0$ & $   0$ & $ 0 $ & $   0$ & $   0$ & $     0$\\
$ 503$ & $   0 $ & $    0$ & $    0 $ & $       0$ & $   28887$ & $  64-021$ & $     0$ & $    0$ & $    0$ & $   0$ & $ 0 $ & $   0$ & $   0$ & $     0$\\
$ 504$ & $3091 $ & $    0$ & $    0 $ & $       0$ & $   28927$ & $       0$  & $     0$ & $    0$ & $    0$ & $   0$ & $ 0 $ & $   0$ & $   0$ & $     0$\\
$ 505$ & $3073 $ & $    0$ & $ 5374 $ & $       0$ & $   28974$ & $ 265-054$ & $     0$ & $    0$ & $    0$ & $   0$ & $ 0 $ & $   0$ & $   0$ & $     0$\\
$ 506$ & $3098 $ & $    0$ & $ 5397 $ & $       0$ & $   29067$ & $ 123-014$ & $     0$ & $    0$ & $    0$ & $   0$ & $ 0 $ & $   0$ & $   0$ & $     0$\\
$ 507$ & $   0 $ & $    0$ & $ 5408 $ & $       0$ & $   29177$ & $ 289-028$ & $     0$ & $    0$ & $    0$ & $   0$ & $ 0 $ & $   0$ & $   0$ & $     0$\\
$ 508$ & $   0 $ & $    0$ & $    0 $ & $       0$ & $   29321$ & $  64-055$ & $     0$ & $    0$ & $    0$ & $   0$ & $ 0 $ & $   0$ & $   0$ & $     0$\\
$ 509$ & $   0 $ & $    0$ & $    0 $ & $       0$ & $       0$ & $       0$ & $     0$ & $    0$ & $    0$ & $   0$ & $ 0 $ & $   0$ & $   0$ &  Segue\_1\\
$ 510$ & $   0 $ & $    0$ & $ 5470 $ & $       0$ & $   29488$ & $  64-073$ & $     0$ & $    0$ & $    0$ & $   0$ & $ 0 $ & $   0$ & $   0$ &  Leo\_I\\
&&&&&&&&&&&&&& \\[-10pt]
\hline
\end{tabular}
\end{table}
\end{landscape}

\section[Locations, distances and redshifts]{Locations, distances and redshifts}
\label{sec:locations}

\subsection{General Procedures}
\label{sec:locations1}
In Tab.~(\ref{tab12}), we list the locations on the sky, distances, redshifts and different names for the galaxies. The galaxies are ordered by ascending right ascension $\alpha(2000.0)$.

Redshift-independent distances are used where available, and the redshift is only used as a distance indicator at distances beyond that of the Fornax Cluster, where the Hubble flow velocities become very large when compared with the random motions of galaxies.

If no redshift, $z$, is known for some galaxy, its morphology and overall structure can be an indicator for whether it is a nearby galaxy or a distant background galaxy. If thereby the membership to some nearby galaxy cluster can be confirmed, the distance of the galaxy can be given fairly accurately as the distance of that galaxy cluster. For galaxies that appear in this catalog, this method has been applied by \citet{Ferguson1989} for confirming the membership of galaxies to the Fornax cluster, by \citet{Misgeld2008} for Hydra Cluster galaxies, by \citet{Misgeld2009} for Centaurus Cluster galaxies, and by \citet{Lieder2012} for Virgo Cluster galaxies. As a result, there are several hundreds of faint dwarf ellipticals in this catalog, for which their membership to a galaxy cluster has been determined by this method, since neither redshifts, nor other estimates for their distances are available for them. However, the procedure of identifying nearby galaxies by their overall appearance seems to be remarkably reliable. There are 64 ETGs that \citet{Ferguson1989} classified as definite or likely members of the Fornax cluster, even though they did not know their $z$, but for which a value for $z$ is now given in the NASA/IPAC Extragalactic Database (NED). For 55 of these galaxies, the value for $z$ now supports their membership to the Fornax cluster. Of the remaining 9 galaxies, \citet{Ferguson1989} considered only two as definite members of the Fornax Cluster, and the remaining 7 only as likely members of the Fornax cluster. Thus, there will be an occasional galaxy that is in this catalogue wrongly assumed to be a faint, nearby ETG, while it is in truth a bright, distant galaxy. However, for the vast majority of the galaxies, the estimate of their distance should be reasonably close to the actual value.

A probable redshift can also be estimated from a given distance via Hubbles Law, which is given as
\begin{equation}
v_{\rm rad}=c \, z_{\rm rad}=H_0 \, d,
\label{eq:hubble}
\end{equation}
where $v_{\rm rad}$ is the recessional velocity of the galaxy from the Sun, $z_{\rm rad}$ is the corresponding redshift, $c$ is the speed of light and $H_0$ is the Hubble constant \citep{Hubble1931}. This has been done here for a large fraction of galaxies that are considered to be members of the Virgo Cluster and the Fornax Cluster, and for most galaxies that are considered to be members of the Hydra Cluster and the Centaurus Cluster, since no redshift has been measured for them. There are also some galaxies in this catalogue that are considered part of the Local Group due to redshift-independent distance estimates, while no redshift has been measured for them. The redshift of those galaxies is set to zero. For galaxies at the distance of the Virgo Cluster or the Fornax cluster, the random motions can be of the order of the velocity of the Hubble flow, so that redshifts calculated with eq~(\ref{eq:hubble}) tell little about their actual heliocentric velocities. We list them nevertheless for completeness and use them for calculating the k-corrections to the magnitudes of those galaxies, in order to be as consistent as possible with the data for more distant galaxies where the k-correction actually matters.

In this catalog, the preferred data for quantifying the motion of the ETGs with respect to the observer are estimates of the radial velocity of the ETGs with respect to the Sun. These values have the advantages that they are the closest to the actual observations and are the correct quantity for performing the k-correction. They are also adequate for determining the membership of a galaxy to a galaxy cluster. Only if such a measurement does not exist for a given galaxy, its $v_{\rm rad}$ is estimated with Eq.~(\ref{eq:hubble}) based on a redshift-independent distance estimate for the galaxy.

Some galaxies have been excluded from this catalogue due to their values for $c \, z_{\rm rad}$ according to NED. Among these galaxies are the 9 galaxies that \citet{Ferguson1989} considered likely or definitive members of the Fornax Cluster, while their redshifts suggests otherwise. They are discarded, because the data on their colours is insufficient to correct their magnitudes properly for redshift (see Section~\ref{sec:magnitudes1} for details on the correction for redshift.) Also, about 20 galaxies from \citet{Scodeggio1998a,Scodeggio1998b} with $c \, z_{\rm rad}>15000$ km/s were discarded. The redshifts of these galaxies are too high to be part of any of the galaxy clusters observed by \citet{Scodeggio1998a, Scodeggio1998b} and cannot be considered galaxies of the nearby Universe, given their redshift.

\subsection{Detailed description of the data}
\label{sec:locations2}

In the following, we describe the entries in Tab.~(\ref{tab12}) column by column.

{\bf Column~1} lists the running number.

{\bf Column~2} lists the right ascension for the epoch J2000.0, $\alpha(2000.0)$.

{\bf Column~3} lists the declination $\delta(2000.0)$.

{\bf Column~4} indicates the origin of the data for $\alpha(2000.0)$ and $\delta(2000.0)$. The entry in this column is set to `1' if the data is taken from the  NASA/IPAC Extragalactic Database (NED) and to `2' if the data stems from \citet{Chilingarian2008}.

{\bf Column~5} shows the distance of the galaxies in Mpc.

{\bf Column~6} indicates the origin of the distance estimate. The number in this column is set to `0' if no data on the distances is available. The number is set to `1' if the given distance estimate is the summary statistics on redshift-independent distance estimates in NED. This is the primary source for $d$. The summary statistics in NED are a quick and convenient way to find redshift-independent distance estimates for a very large number of galaxies. The disadvantage of those summary statistics is that they are based on data from original sources, without any homogenisation or correction for systematic errors. The advantage of the values coming from NED is however, that at least for large galaxies or close galaxies, the distance has been estimated more than once. The value quoted from NED is then the mean of all those individual estimates, so that systematic errors tend to cancel out. At least for galaxies with several distance estimates considered for the summary statistics in NED, the resulting distance estimates should therefore be quite reliable. The number is set to `2' if the distance estimate comes from \citet{Cappellari2013}. The number is set to `3' if the distance estimate comes from \citet{Toloba2012}. The number is set to `4' if the distance estimate comes from \citet{Guerou2015}. If the entry in column~6 is none of previous numbers, no redshift-independent distance estimate was available for the respective galaxies. For those galaxies, the number is set to `101' if $d$ was calculated directly with Eq.~\ref{eq:hubble} from the $z$ of the galaxy. The number is set to `102' if the galaxy is considered a member of the Virgo Cluster, and the estimate for its distance is based on that membership. The distance of the galaxy is then considered to be 16.5 Mpc, which is the distance of the Virgo Cluster according to \citet{Mei2007}. The number is set to `103' if it is considered to be a member one of the galaxy clusters NGC~383, NGC~507, Abell~262, Cancer, Abell~1367, Coma, Pegasus or Abell~2634. The heliocentric velocity, $v_{\rm hel}$ of those galaxy clusters is given in \citet{Scodeggio1998a}, which is used here to estimate the distance of those galaxy clusters with Eq.~(\ref{eq:hubble}), and which is in turn the distance that we assign to the galaxies that we consider members of those galaxy clusters. The galaxies that we consider to be members of these clusters are those that lie in the direction of the galaxy clusters in question and have heliocentric velocities $\pm 2000 \, {\rm km/s}$ of the heliocentric velocity of the appropriate galaxy cluster. The number is set to `104' in column~6 for galaxies that \citet{Chilingarian2008} consider members of the galaxy cluster Abell~496, with a distance of 133 Mpc. The membership of all these galaxies to Abell~496 is confirmed by their redshift. The number in column~6 is set to `105' if the galaxy is considered to be a member of the Fornax Galaxy Cluster, and the estimate for its distance is based on that membership. The distance to the Fornax Cluster considered to be 18.967 Mpc (cf. \citealt{Freedman2001}). The number in column~6 is set to `106' if the galaxy is considered to be a member of the Hydra Galaxy Cluster, and the estimate for its distance is based on that membership. The distance to the Hydra Cluster is considered to be 47.206 Mpc (cf. \citealt{Mieske2005}). The number in column~6 is set to `107' if the galaxy is considered to be a member of the Centaurus Galaxy Cluster, and the estimate for its distance is based on that membership. The distance to the Centaurus Cluster is considered to be 45.290 Mpc (cf. \citealt{Mieske2005}).

{\bf Column~7} lists the number of redshift-independent estimates from which the mean of the redshift-independent estimates quoted in this catalogue was calculated. This number can be as large as 100 for redshift-independent distances quoted from NED. It is 0, if the distance is estimated from the assignment of a galaxy to a galaxy cluster or from its redshift.

{\bf Column~8} lists the radial velocities of the galaxies, $v_{\rm rad}$. Heliocentric velocities based on observed redshifts of individual galaxies are preferred over other estimates for $v_{\rm rad}$.

{\bf Column~9} lists the source for the radial velocities of the galaxies. The number in column~9 is set to `1' if the source for $v_{\rm rad}$ is NED. This is the primary source for $v_{\rm rad}$ in this catalogue. The number in Column~9 is set to `2' if the source for $v_{\rm rad}$ is \citet{Kourkchi2012a}, to `3' if the source for $v_{\rm rad}$ is \citet{Chilingarian2008}, to `4' if the source is HyperLeda, and to `5' if the source is \citet{Scodeggio1998a}. All of the previous numbers indicate heliocentric velocities based on the observed redshifts of the respective galaxies. If the number in column~9 is none of the previous numbers, no redshift has directly been measured for the given galaxy. The number in column~9 is set to `101' if $v_{\rm rad}$ is calculated from Eq.~(\ref{eq:hubble}) from a redshift-independent estimate of $d$. The number in column~9 is set to `102' if the galaxy is considered to be a member of the Local Group, and its value for $v_{\rm rad}$ has been set to 0 km/s. The number in column~9 is set to `103' if the galaxy is considered to be a member of the Virgo Cluster, and its value for $v_{\rm rad}$ has been calculated with Eq.~(\ref{eq:hubble}) using the distance to the Virgo Cluster. The distance to the Virgo Cluster is 16. 5 Mpc according to \citet{Mei2007}, leading to $v_{\rm rad} =1172$ km/s. The number in column~9 is set to `104' if the galaxy is considered to be a member of the Fornax Cluster, and its value for $v_{\rm rad}$ has been calculated with Eq.~(\ref{eq:hubble}) using the distance to the Fornax Cluster. The distance to the Fornax Cluster is 18.967 Mpc according to \citet{Freedman2001}, leading to $v_{\rm rad} =1347$ km/s. The number in column~9 is set to `105' if the galaxy is considered to be a member of the Hydra Cluster, and its value for $v_{\rm rad}$ has been calculated with Eq.~(\ref{eq:hubble}) using the distance to the Hydra Cluster. The distance to the Hydra Cluster is 47.206 Mpc according to \citet{Mieske2005}, leading to $v_{\rm rad} =3352$ km/s.  The number in column~9 is set to `106' if the galaxy is considered to be a member of the Centaurus Cluster, and its value for $v_{\rm rad}$ has been calculated with Eq.~(\ref{eq:hubble}) using the distance to the Centaurus Cluster. The distance to the Centaurus Cluster is 45.290 Mpc according to \citet{Mieske2005}, leading to $v_{\rm rad} =3216$ km/s. A '107' indicates galaxies in the Local Group with unknown $v_{\rm rad}$. Their $v_{\rm rad}$ has been set to 0 km/s here.

\begin{table*}
\caption{\label{tab12} Locations, distances and redshifts of the galaxies in this catalogue. A portion of the Table is shown here for guidance regarding its contents and form. A detailed description of the contents of this table is given in Section~(\ref{sec:names2}). A detailed description of the data in the table is given in section~(\ref{sec:locations2}). The table in its entirety can be downloaded at http://www.astro-udec.cl/mf/catalogue/}
\centering
\vspace{2mm}
\begin{tabular}{rrrrrrrrr}
\hline
&&&&&&&& \\[-10pt]
id       & $\alpha(2000.0) $ & $\delta(2000.0)$ & s. & $ d $   & $ N_d $& s. & $v_{\rm rad}$ & s. \\
          &  $[h:m:s]$                 &  $[d:m:s]$                &             & $[{\rm Mpc}]$    &              &            &  $[{\rm km/s}]$             &             \\  
\hline
$ 501$ & $09:45:53.4$ & $-00:16:06$ & $ 1$ & $ 27.200$ & $   1$ & $   2$ & $  1853$ & $   1$\\
$ 502$ & $09:52:08.1$ & $ 29:14:10$ & $ 1$ & $ 26.217$ & $   6$ & $   1$ & $  1562$ & $   1$\\
$ 503$ & $09:59:43.5$ & $ 11:39:39$ & $ 1$ & $ 41.000$ & $   1$ & $   2$ & $  2833$ & $   1$\\
$ 504$ & $10:00:14.3$ & $-19:38:13$ & $ 1$ & $ 52.525$ & $   4$ & $   1$ & $  3964$ & $   1$\\
$ 505$ & $10:00:52.1$ & $ 55:37:08$ & $ 1$ & $ 24.520$ & $   5$ & $   1$ & $  1173$ & $   1$\\
$ 506$ & $10:02:16.7$ & $ 24:42:40$ & $ 1$ & $ 20.267$ & $   3$ & $   1$ & $  1397$ & $   1$\\
$ 507$ & $10:03:51.9$ & $ 59:26:10$ & $ 1$ & $ 45.800$ & $   1$ & $   2$ & $  2989$ & $   1$\\
$ 508$ & $10:05:51.2$ & $ 12:57:41$ & $ 1$ & $ 40.900$ & $   1$ & $   2$ & $  2816$ & $   1$\\
$ 509$ & $10:07:04.0$ & $ 16:04:55$ & $ 1$ & $  0.023$ & $   2$ & $   1$ & $   208$ & $   1$\\
$ 510$ & $10:08:28.1$ & $ 12:18:23$ & $ 1$ & $  0.243$ & $  28$ & $   1$ & $   285$ & $   1$\\
&&&&&&&& \\[-10pt]
\hline
\end{tabular}
\end{table*}

\section[Magnitudes]{Magnitudes and luminosities}
\label{sec:magnitudes}

\subsection{General procedures}
\label{sec:magnitudes1}

In Tables~(\ref{tab21}) and~(\ref{tab22}), we list the magnitudes of the galaxies in the $ugriz$-system that is used by the Sloan Digital Sky Survey (SDSS), as well as in the standard $UBVR_{C}I_{C}$-system (i.e. the Johnson system for $U$ $B$ and $V$ and the Cousins system for $R$ and $I$). In practise, we do not list $R_{C}$, since we hardly found any data on this magnitude in the literature we were using for this catalogue. For a detailed treatise of these photometric systems, we refer the reader to \citet{Bessell2005}. If these magnitudes are available, they are combined with the distances in Table \ref{tab12} in order to calculate the luminosities of the galaxies in Solar units. This requires values for the absolute magnitude of the Sun, which are taken from \citet{Binney1998} for the standard system and from \citet{Blanton2007} for the SDSS-system. The resulting luminosities are presented in Tables~(\ref{tab31}) and~(\ref{tab32}).

\subsubsection{Corrections for extinction and redshift}
\label{sec:extinction}

The apparent magnitudes presented in this catalogue are corrected for the effects of extinction and redshift wherever possible (see Section~\ref{sec:magnitudes2} for details), if this has not already been done for the original data. The correction is performed in two steps. 

The first step is the correction for extinction. Magnitudes, luminosities and colours are already corrected for internal and galactic foreground extinction, if the source for the data is HyperLeda, or the catalogue on nearby galaxies by \citet{Karachentsev2013}. Data on magnitudes, luminosities and colours from any other source are only corrected for galactic foreground extinction, using the data on this given in NED, which are based on the dust map by \citet{Schlegel1998} and the recalibration of this dust map by \citet{Schlafly2011}. Neglecting a possible internal extinction seems however justifiable, since most galaxies have no internal extinction according to the data given in HyperLeda or \citet{Karachentsev2013}. This first step of correction is performed for every galaxy in this catalogue.

The second step is the correction for redshift. This is done using analytic approximations for the $k$-correction provided by \citet{Chilingarian2010}, which are functions of colours (corrected for extinction) and redshift. A meaningful application of this correction is possible for almost all galaxies in this catalogue, except for the galaxies in the Fornax Cluster and the galaxies in the Local group, where the available data is insufficient. Correcting the magnitudes of these galaxies for redshifts is, however, more a formality than a necessity. Doing so would increase the homogeneity of the catalog, but the redshifts of galaxies in the Fornax Cluster or the Local Group are so small that correcting them for redshifts hardly matters for all practical purposes.

\subsubsection{Gauging magnitudes from different sources to a common standard}
\label{sec:gaugeMagnitudes}

Despite the corrections for extinction and redshift, there are often quite significant systematic offsets between data on the apparent magnitude of the same galaxy in the same passband, but from different sources for the data. In order to homogenise our data on the apparent magnitudes of galaxies (and related quantities) from different sources, we quantify the systematic offset between that data in different samples, if there is sufficient overlap between the two samples.

First, we choose one source as the primary source. For each galaxy for which data on the apparent magnitude are available in the primary source as well as the secondary source, the difference between the two estimates for the same quantity is
\begin{equation}
\Delta m_X^i=m_{X\, {\rm prim}}^{i}-m_{X\, {\rm sec}}^{i},
\label{eq:gaugelum}
\end{equation}
where $m_{X\, {\rm prim}}^{i}$ is the apparent magnitude of the $i$th galaxy in the passband $X$ according to the primary source and $m_{X\, {\rm sec}}^{i}$ is the same quantity according to the secondary source. The adopted value for $m_{X}$ of the $i$th galaxy in the secondary sample are calculated as
\begin{equation}
m_{X \, {\rm calc}}^{i}=m_{X\, {\rm sec}}^{i}+\Delta m_X, 
\end{equation}
where $\Delta m_X$ is the median of the $i$ different values for $\Delta m_X^i$ calculated with Eq.~(\ref{eq:gaugelum}).

\subsubsection{Estimating magnitudes using correlations between observed values}
\label{sec:lumcorrelation}

For many ETGs in this catalog, apparent magnitudes have been measured in the standard system, but not in the SDSS-system, or vice versa. However, there are also many galaxies for which data on the apparent magnitudes have been measured in both photometric systems. This data can be used for estimating apparent magnitudes that have not been observed. For this, correlations between the luminosities of the galaxies in the standard system and the luminosities of the same galaxies in the SDSS system are used.

We use the data on the apparent magnitudes of the ETGs after correcting them for extinction and redshift (see Section~{\ref{sec:extinction}) and gauging them to a common standard (see Section~\ref{sec:gaugeMagnitudes}). These apparent magnitudes are transformed into luminosities in Solar units, using the distances given in Tab.~(\ref{tab12}) and the absolute magnitudes of the Sun given in Tables~(\ref{tab:LumfitSDSS} and~(\ref{tab:LumfitStandard}). These luminosities in both photometric systems are used to calibrate linear relations in log-log space between them. These relations express the average luminosities in the standard system as functions of the luminosities in the SDSS-system, and vice versa. Thus, estimates for the luminosity of a galaxy can be calculated with equations of the form
\begin{equation}
\log_{10} \left(\frac{L_Y}{10^6 \, L_{Y\, \odot}}\right)=a\log_{10} \left(\frac{L_X}{10^6 \, L_{X\, \odot}}\right)+b,
\label{eq:luminosityfit}
\end{equation}
where $L_X$ and $L_Y$ are luminosities the passbands $X$ and $Y$ and $a$ and $b$ are the best-fitting parameters obtained in a least-squares fit. If the uncertainty of $\log_{10}(L_X)$ or $\log_{10}(L_Y)$ is larger than 0.10 for a given galaxy, it is not considered for finding the best-fitting parameters of Eq.~(\ref{eq:luminosityfit}). The same is true if $L_{X}< 10^7 \ {\rm L}_{\odot}$ or  $L_{Y}< 10^7 \ {\rm L}_{\odot}$, since estimates for $L_X$ and $L_Y$ become rather uncertain at the lowest luminosities. The considered combinations of luminosities and the resulting best-fitting parameters $a$ and $b$ are listed in Tab.~\ref{tab:LumfitSDSS} for transformations from the SDSS-system into the standard-system, and in Tab.~\ref{tab:LumfitStandard} for transformations from the standard-system into the SDSS-system.

There are many ETGs for which $L_Y$ can be calculated from more than one $L_X$. We therefore calculate the final estimate for $L_Y$ as  
\begin{equation}
L_{Y\, \rm calc}=\frac{1}{S} \, \sum_{i=1}^{S} \log_{10}\left(\frac{L_{Y}^{i}}{L_{Y\, \odot}}\right),
\label{eq:meanLuminosity} 
\end{equation}
where $L_{Y}^{i}$ are the estimates for the luminosity $L_Y$ and $S$ is the number of such individual estimates for $L_Y$, with $S \le 2$ in practice. This summation over the different $L_{Y}^{i}$ is done in order to attenuate the influence on the best estimate for $L_Y$ from occasionally flawed measurements of $L_X$ in single passbands.

The uncertainty for the entries in Tables~(\ref{tab31}) and~(\ref{tab32}) based on Eqs.~(\ref{eq:luminosityfit}) and~(\ref{eq:meanLuminosity}) is estimated using the galaxies for which observed luminosities in the standard system as well as observed luminosities in the SDSS-system are available. For this purpose, $L_{Y\, {\rm calc}}$ is also estimated for those galaxies using Eqs.~(\ref{eq:luminosityfit}) and~(\ref{eq:meanLuminosity}). The uncertainty for $L_{Y\, {\rm calc}}$ is then estimated using
\begin{equation}
\sigma^2=\frac{1}{N-1} \, \sum^{N}_{i=1}(L_{Y\, {\rm calc}}^{i}-L_{Y\, {\rm obs}}^{i})^2,
\label{eq:sigmalum}
\end{equation}
where $\sigma$ is the standard deviation, $L_{Y\, {\rm obs}}^{i}$ is the logarithm to the base of 10 of the observed luminosity of the $i$th galaxy in the $Y$-passband, $L_{Y\, {\rm calc}}^{i}$ is the value for the $Y$-band luminosity calculated from Eqs.~(\ref{eq:luminosityfit}) and~(\ref{eq:meanLuminosity}) for the same galaxy and $N$ is the number of all galaxies for which $L_{Y\, {\rm calc}}^{i}$, as well as $L_{Y\, {\rm obs}}^{i}$ are available. Thus, the uncertainty is estimated from the scatter around the relation
\begin{equation}
L_{Y\, {\rm calc}}=L_{Y\, {\rm obs}},
\label{eq:compLuminosity}
\end{equation}
where $L_{Y\, {\rm obs}}$ is the observed luminosity in the $Y$-band and $L_{Y\, {\rm calc}}$ the value for the $Y$-band luminosity calculated from Eqs.~(\ref{eq:luminosityfit}) and~(\ref{eq:meanLuminosity}).

Fig.~(\ref{fig:LumScatter}) illustrates how well the above procedure can reproduce the observed Johnson $B$-band luminosities from SDSS $u$ and $g$-band luminosities. For comparison, we also show the results from using $B = u-1.0286\, [0.79811\, (u-g)-1.2638]$, which is a conversion formula introduced in \citet{Blanton2007}. We find that with our fit, the data are better centred on a line indicating equality between observed and calculated values than with the relation from \citet{Blanton2007}. This is no surprise, because in the first case, we use relations on the very same data to which they were fitted in the first place. The scatter of the data seems similar in either case. This indicates that our method, using a mean of single passband transformations, is not less precise than the more conventional method by \citet{Blanton2007}, using a magnitude and a colour. However, the advantage of our method is that it can also be used if only input magnitudes from two different sources are available, i.e. where no meaningful colour can be estimated, and that it can be adapted to cases where only one input magnitude is available.

\begin{figure*}
\centering
\includegraphics[scale=0.78]{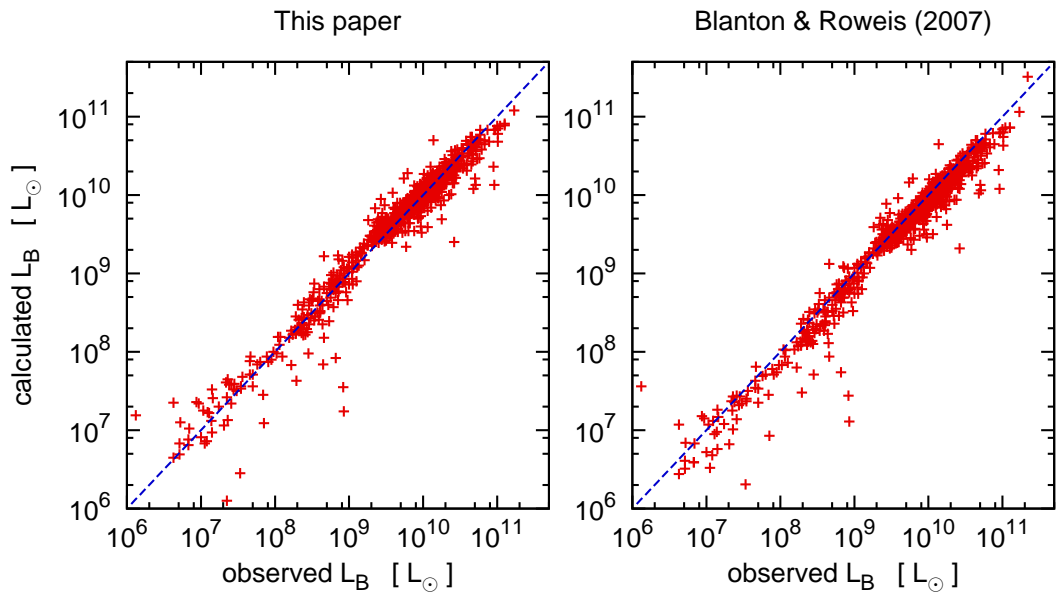}
\caption{\label{fig:LumScatter} The estimates for $L_B$ against the observed $L_B$ for the galaxies for which both estimates are available. The values are calculated according to the procedure described in Section~(\ref{sec:magnitudes1}) in the left panel and with with a relation from Table~2 in \citet{Blanton2007} in the right panel. The diagonal line indicates equality between both quantities. Figures like this one, but for other passbands, show the same characteristics.}
\end{figure*}

Using the distances listed in Tab.~(\ref{tab12}) and the absolute magnitudes of the Sun in the different passbands, the estimates for the luminosity obtained with Eqs.~(\ref{eq:luminosityfit}) and~(\ref{eq:meanLuminosity}) are converted into apparent magnitudes. The corresponding uncertainties calculated with Eq.~(\ref{eq:sigmalum}) are converted into uncertainties of the apparent magnitude by multiplying them by 2.5. The results are listed in Tables~(\ref{tab21}) and~(\ref{tab22}) for galaxies without measured magnitudes in the according passbands.

For apparent magnitudes in the standard system that have been estimated from SDSS-luminosities with Eqs.~(\ref{eq:luminosityfit}) and~(\ref{eq:meanLuminosity}), the provenience of these estimates is indicated in Tables~(\ref{tab21}) and~(\ref{tab31}) by numbers that start with `10' followed by two digits that specify the luminosities in the SDSS-system that entered the estimate. In this context, $u=1$, $g=2$, $r=3$, $i=4$ and $z=5$. If only a single passband was used, the other digit is set to `0'. The numbers indicating how magnitudes in the SDSS system were calculated from the standard system follow the same principle. However, in this context, $U=1$, $B=2$, $V=3$ and $I=4$.

\begin{table*}
\caption{\label{tab:LumfitSDSS} Parameters for linear fits of the logarithmic luminosities from the SDSS-system versus logarithmic luminosities from the standard system. The first column lists the passbands in the standard system that are to be estimated via Eq.~(\ref{eq:luminosityfit}). The second column lists the absolute magnitude of the Sun in the passbands given in Column~1. The third and the fourth Column give the best-fitting parameters in Eq.~(\ref{eq:luminosityfit}), if the passband in the the standard system is to be estimated from the $u$-band luminosity in the SDSS-system. The fifth column lists the number of galaxies that were used for calculating the parameters listed in columns~3 and~4. In the next columns, the scheme of columns 3 to 5 is repeated for the $g$-band, the $r$-band, the $i$-band and the $z$-band of the SDSS-system.}
\centering
\vspace{2mm}
\begin{tabular}{lrrrrrrrrrrrrrrrr}
\hline
&&&&&&&&&&&&&&&\\[-10pt]
band  & $M_{X}$ &  $a_u$   & $b_u$    & $N_u$ & $a_g$   &  $b_g$   & $N_g$ &  $a_r$   &  $b_r$   & $N_r$ &  $a_i$   &  $b_i$   & $N_i$ &  $a_z$   &  $b_z$   & $N_z$ \\
\hline
U     & 5.61        &  $0.999$ &  $0.199$ & $327$ & $0.934$ &  $0.230$ & $330$ &  $-$        &  $-$         & $-$       &  $-$         &  $-$         & $-$      &  $-$         &  $-$        & $-$   \\
B     & 5.48        &  $1.022$ &  $0.214$ & $851$ & $0.946$ &  $0.293$ & $860$ &  $-$        &  $-$         & $-$       &  $-$         &  $-$         & $-$      &  $-$         &  $-$        & $-$   \\
V     & 4.83        &  $-$         & $-$         & $-$       & $1.008$ &  $0.155$ & $403$ &  $0.974$ &  $0.177$ & $404$  &  $-$        & $-$          & $-$      &  $-$         &  $-$        & $-$   \\
I     & 4.08         &  $-$         &  $-$         & $-$      & $-$         &  $-$         & $-$      &  $-$        &  $-$         & $-$       &  $0.959$ &  $0.178$ & $158$ &  $0.932$ &  $0.173$ & $143$ \\

&&&&&&&&&&&&&&&\\[-10pt]
\hline
\end{tabular}
\end{table*}

\begin{table*}
\caption{\label{tab:LumfitStandard} Linear fits of the logarithmic luminosities from the standard-system versus logarithmic luminosities from the SDSS-system. The first column lists the passbands in the SDSS-system that are to be estimated via Eq.~(\ref{eq:luminosityfit}). The second column lists the absolute magnitude of the Sun in the passbands given in Column~1. The third and the fourth Column give the best-fitting parameters in Eq.~(\ref{eq:luminosityfit}), if the passband in the SDSS-system is to be estimated from the $U$-band luminosity in the standard-system. The fifth column lists the number of galaxies that were used for calculating the parameters listed in columns~3 and~4. In the next columns, the scheme of columns 3 to 5 is repeated for the $B$-band, the $V$-band and the $I$-band of the standard-system.}
\centering
\vspace{2mm}
\begin{tabular}{lrrrrrrrrrrrrr}
\hline
&&&&&&&&&&&&&\\[-10pt]
band  & $M_{X}$ &  $a_U$   & $b_U$    & $N_U$ &  $a_B$   &  $b_B$    & $N_B$ &  $a_V$   &  $b_V$     & $N_V$ &  $a_I$     &  $b_I$     & $N_I$ \\
\hline
u     & 6.38        &  $0.945$   &  $0.004$ & $327$  &  $0.919$ &  $0.080$ & $851$ &  $-$         &  $-$          & $-$       &  $-$         &  $-$         & $-$ \\
g     & 5.12        &  $1.008$   & $-0.023$ & $331$  &  $1.007$ & $-0.124$ & $860$ &  $-$         &  $-$          & $-$       &  $-$         & $-$          & $-$ \\
r     & 4.64         &  $-$           & $-$          & $-$       &  $-$        & $-$          & $-$      &  $0.992$ &  $-0.057$ & $404$  &  $1.017$ & $-0.203$ & $134$ \\
i     & 4.53         &  $-$           &  $-$         & $-$       &  $-$        & $-$          & $-$      &  $-$         &  $-$          & $-$       &  $1.013$ & $-0.099$ & $158$ \\
z     & 4.51        &  $-$           &  $-$         & $-$       &  $-$        &  $-$         & $-$      &  $-$         &  $-$          & $-$       &  $1.046$ & $-0.102$ & $144$ \\

&&&&&&&&&&&&&\\[-10pt]
\hline
\end{tabular}
\end{table*}

\subsection{Detailed description of the data}
\label{sec:magnitudes2}

\subsubsection{Magnitudes in the standard system}
\label{sec:magnitudes21}

Below, the entries in Tab.~(\ref{tab21}) are described column by column. The entries for magnitudes and their uncertainties in that table are set to $-9.99$ if a value is not available, and always positive otherwise. The magnitudes are corrected for extinction and redshift. Uncertainties, if available, always come from the same source as the value for the magnitude itself.

{\bf Column~1} lists the running number of the galaxy.

{\bf Column~2} lists the apparent $U$-band magnitude, $m_U$, of the galaxy. 

{\bf Column~3} quotes the uncertainty of $m_U$.

{\bf Column~4} lists the source for the quoted value for $m_U$. The number is set to `0' if no data on $m_U$ is available. The number in that column is set to `1' if $m_U$ is calculated from the apparent $B$-band magnitude from HyperLeda and the $(U-B)$ colour from the same source. This is the primary source for $m_U$ in this catalog, since HyperLeda provides for a vast number of galaxies of all sizes apparent $B$-band magnitudes and $(U-B)$-colours that have already been homogenised and corrected for redshift and extinction. The apparent $U$-band magnitudes according to HyperLeda thus provide the standard to which the values for $m_U$ from other sources are gauged. A `2' indicates that $m_U$ comes from \citet{Eisenhardt2007}, who provide photometry for galaxies in the Coma Cluster. Using the method described in Section~(\ref{sec:gaugeMagnitudes}), we gauge the values for $m_{U}$ of these galaxies by $\Delta m_U=-0.19$, where the estimate for $\Delta m_U$ is based on 33 ETGs. A number of the form `10xx' indicates that the value for $m_U$ has been calculated from the $U$-band luminosities estimated with Eq.~(\ref{eq:luminosityfit}), using the distances in Tab.~(\ref{tab12}) and the absolute magnitude of the Sun in Tab.~(\ref{tab:LumfitSDSS}). The meaning of those numbers is specified at the end of Section~(\ref{sec:lumcorrelation}).
 
{\bf Column~5} lists the apparent $B$-band magnitude, $m_B$, of the galaxy. 

{\bf Column~6} quotes the uncertainty of $m_B$.

{\bf Column~7} lists the sources for the quoted values for $m_B$. The number in this column is set to `0' if no data on $m_B$ is available. The number is set to `1' if the source for $m_B$ is HyperLeda. This is the primary source for $m_B$ in this catalog, to which values for $m_B$ from other sources are gauged if possible. The reasons for this choice are the same as for the $U$-band. A `2' indicates that $m_B$ comes from \citet{Eisenhardt2007}, gauged by $\Delta m_B=-0.52$, based on 77 ETGs. A `3' indicates that $m_B$ comes from \citet{Karick2003}, gauged by $\Delta m_B=-0.38$, based on 121 ETGs. A `4' indicates that the galaxy is a member of the Local Group whose value for $m_B$ comes from the catalogue by \citet{Karachentsev2013}, gauged by $\Delta m_B=-0.005$, based on 28 ETGs. A `5' indicates that $m_B$ comes from \citet{Koleva2011}. These values are not gauged to the standard from HyperLeda, since the overlap between the data from \citet{Koleva2011} and the data from HyperLeda is too small. A number of the form `10xx' indicates that the value for $m_B$ has been calculated as described in Section~(\ref{sec:lumcorrelation}).

{\bf Column~8} lists the apparent $V$-band magnitude, $m_V$, of the galaxy.

{\bf Column~9} quotes the uncertainty of $m_V$.

{\bf Column~10} lists the source for the quoted values for $m_V$. The number in this column is set to `0' if no data on $m_V$ is available. It is set to `1' if $m_V$ is calculated from the apparent $B$-band magnitude in HyperLeda and the $(B-V)$ colour from the same source. Again, this is the primary source for $m_V$ in this catalog, to which values for $m_V$ from other sources are gauged, if possible. The reasons for this choice are the same as for the $U$-band. A `2' indicates that $m_V$ comes from \citet{Toloba2012}. These values are not gauged to the standard from HyperLeda, since the overlap between the data from HyperLeda and the data from \citet{Toloba2012} is too small. A `3' indicates that $m_V$ comes from \citet{Geha2003}. These values are not gauged for the same reason. A `4' indicates that $m_V$ is taken from \citet{Eisenhardt2007}, gauged by $\Delta m_V=-0.43$, based on 45 ETGs. A `5' indicates that $m_V$ comes from \citet{Karick2003}. There are only 13 galaxies, for which values for $m_V$ are available also from HyperLeda, which might be too little to calculate a meaningful median from the results from Eq.~(\ref{eq:gaugelum}). We therefore gauge the values for $m_V$ from \citet{Karick2003} to the values for $m_V$ from HyperLeda by assuming $\Delta m_V^i=-0.38$ for them, i.e. the values for $m_V$ from \citet{Karick2003} are shifted by the same amount as the values for $m_B$ from \citet{Karick2003}. This preserves the $(B-V)$-colours that are calculated from \citet{Karick2003}, and which seem realistic. A `6' indicates that the source for $m_V$ is \citet{Mieske2007}. There is not much overlap between the data on $m_V$ from \citet{Mieske2007} and the data on $m_V$ from HyperLeda, but there is some overlap between the data on $m_V$ from \citet{Mieske2007} and the data on $m_V$ from \citet{Karick2003}. Using our method, where we consider the data from \citet{Karick2003} as the primary sample and the data from \citet{Mieske2007} as the secondary sample, we gauge the values for $m_{V}$ from \citet{Mieske2007} by $\Delta m_V=-0.07$, where the estimate for $\Delta m_V$ is based on 22 ETGs. Noting that the offset of the data on $m_V$ from \citet{Karick2003} from the data from HyperLeda is assumed to be $\Delta m_V=-0.38$, we use $\Delta m_V=-0.45$ for gauging the data on $m_V$ from \citet{Mieske2007} to our standard. Data on $m_V$ from \citet{Karick2003} are preferred over those from \citet{Mieske2007} in this catalog, because of the overlap between the data on $m_B$ from \citet{Karick2003} and HyperLeda, which makes it possible to gauge the data from \citet{Karick2003} directly to our standard at least in the $B$-band. A `7' indicates that the galaxy is a member of the Local Group, whose value for $m_V$ comes from the catalogue by \citet{McConnachie2012}. These values have generally already been corrected for Galactic foreground extinction \citep{McConnachie2012}. A `8' indicates that the galaxy is a member of the Virgo Cluster whose value for $m_V$ comes from \citet{Lieder2012}. A `9' indicates that the galaxy is a member of the Hydra Cluster whose extinction-corrected value for $m_V$ comes from \citet{Misgeld2008}. A `10' in column~10 indicates that the galaxy is a member of the Centaurus Cluster, whose extinction-corrected value for $m_V$ comes from \citet{Misgeld2009}. The last four datasets have no significant overlap with any of the other datasets used here, and are therefore not subject to gauging. A number of the form `10xx' indicates that the value for $m_V$ has been calculated as described in Section~(\ref{sec:lumcorrelation}).

{\bf Column~11} lists the apparent $I$-band magnitude, $m_I$, of the galaxy. 

{\bf Column~12} quotes the uncertainty of $m_I$.

{\bf Column~13} lists the source for the quoted values for $m_I$. The number in this column is set to `0' if no data on $m_I$ is available. It is set to `2' if $m_I$ is taken from \citet{Eisenhardt2007}. There is no data on $m_I$ in the standard system in HyperLeda. We therefore gauge the values for $m_I$ from \citet{Eisenhardt2007} by assuming $\Delta m_I=-0.43$ for them, i.e. the values for $m_I$ from \citet{Eisenhardt2007} are gauged by the same amount as the values for $m_V$ from \citet{Eisenhardt2007}. This preserves the $(V-I)$ colours calculated from $m_V$ and $m_I$ from \citet{Eisenhardt2007}, which seem realistic if no modification is applied to them. There is no overlap between the sample on $m_I$ from \citet{Eisenhardt2007} with any other samples on $m_I$ used in this catalogue. The number is set to `3' if $m_I$ comes from \citet{Karick2003}. We add $\Delta m_I=-0.38$ to these values. We thereby shift the respective values for $m_I$ by the same amount with respect to the original data as we shift the original values for $m_V$ from \citet{Karick2003}. This preserves $(V-I)$ colours calculated from $m_V$ and $m_I$ from \citet{Karick2003}, which seem realistic if no modification is applied to them. The number is set to `4' if $m_I$ comes from \citet{Mieske2007}. There is some overlap between the data on $m_I$ from \citet{Mieske2007} and the data on $m_I$ from \citet{Karick2003}, which can be used to homogenise this data. Using our method, we consider the data from \citet{Karick2003} as the primary sample and the data from \citet{Mieske2007} as the secondary sample, and gauge the values for $m_{I}$ from \citet{Mieske2007} by $\Delta m_V=-0.15$, based on 23 ETGs. Noting that the data on $m_I$ from \citet{Karick2003} are already shifted by $\Delta m_I=-0.38$, we use $\Delta m_I=-0.53$ for gauging the data on $m_I$ from \citet{Mieske2007} to our standard. As for $m_V$, data from \citet{Karick2003} on $m_I$ are preferred over data from \citet{Mieske2007}. A `5' indicates that the galaxy is a member of the Virgo Cluster, whose value for $m_I$ is taken from \citet{Lieder2012}. A `6' indicates that the galaxy is a member of the Hydra Cluster, whose value for $m_I$ is calculated from the extinction-corrected $m_V$ and $(V-I)$-colours in \citet{Misgeld2008}. A `7' indicates that the galaxy is a member of the Centaurus Cluster whose value for $m_I$ is calculated from the extinction-corrected $m_V$ and the $(V-I)$-colours in \citet{Misgeld2009}. The last three datasets have no significant overlap with any of the other datasets used here, and are therefore not subject to gauging. A number of the form `10xx' indicates that the value for $m_I$ has been calculated as described in Section~(\ref{sec:lumcorrelation}).
 
\begin{table*}
\caption{\label{tab21} Standard-system magnitudes of the galaxies in this catalogue. A portion of the Table is shown here for guidance regarding its contents and form. A detailed description of the contents of this table is given in Section~(\ref{sec:magnitudes2}). The table in its entirety can be downloaded at http://www.astro-udec.cl/mf/catalogue/}
\centering
\vspace{2mm}
\begin{tabular}{rrrrrrrrrrrrr}
\hline
&&&&&&&&&&&& \\[-10pt]
id     & $ m_U$ & $ dm_U$    & s.   & $ m_B$  & $ dm_B$  & s.   & $ m_V $  & $ dm_V$  & s.   & $ m_I $  & $dm_I$   & s.  \\
       & $[{\rm mag}]$   & $[{\rm mag}]$       &               & $[{\rm mag}]$   & $[{\rm mag}]$     &               &$[{\rm mag}]$ &  $[{\rm mag}]$    &               &  $[{\rm mag}]$ &  $[{\rm mag}]$ &              \\
\hline
$ 501$ & $ 14.08$ & $  0.50$ & $  1012$ & $ 13.75$ & $ -9.99$ & $     1$ & $ 12.84$ & $  0.47$ & $  1023$ & $ 11.97$ & $  0.45$ & $  1045$\\
$ 502$ & $ 13.06$ & $ -9.99$ & $     1$ & $ 12.97$ & $ -9.99$ & $     1$ & $ 12.34$ & $ -9.99$ & $     1$ & $ 12.24$ & $  0.45$ & $  1045$\\
$ 503$ & $ 14.80$ & $  0.50$ & $  1012$ & $ 14.63$ & $ -9.99$ & $     1$ & $ 13.58$ & $  0.47$ & $  1023$ & $ 12.78$ & $  0.45$ & $  1045$\\
$ 504$ & $ -9.99$ & $ -9.99$ & $     0$ & $ 11.88$ & $ -9.99$ & $     1$ & $ 10.96$ & $ -9.99$ & $     1$ & $ -9.99$ & $ -9.99$ & $     0$\\
$ 505$ & $ 14.68$ & $ -9.99$ & $     1$ & $ 14.04$ & $ -9.99$ & $     1$ & $ 13.40$ & $ -9.99$ & $     1$ & $ 12.44$ & $  0.45$ & $  1045$\\
$ 506$ & $ 13.46$ & $ -9.99$ & $     1$ & $ 12.68$ & $ -9.99$ & $     1$ & $ 11.90$ & $ -9.99$ & $     1$ & $ 11.07$ & $  0.45$ & $  1045$\\
$ 507$ & $ 14.39$ & $ -9.99$ & $     1$ & $ 14.65$ & $ -9.99$ & $     1$ & $ 14.12$ & $ -9.99$ & $     1$ & $ 13.13$ & $  0.45$ & $  1045$\\
$ 508$ & $ 15.22$ & $  0.50$ & $  1012$ & $ 14.98$ & $ -9.99$ & $     1$ & $ 14.06$ & $  0.47$ & $  1023$ & $ 13.21$ & $  0.45$ & $  1045$\\
$ 509$ & $ -9.99$ & $ -9.99$ & $     0$ & $ 14.41$ & $ -9.99$ & $     4$ & $ 13.30$ & $  0.50$ & $     4$ & $ -9.99$ & $ -9.99$ & $     0$\\
$ 510$ & $  8.56$ & $ -9.99$ & $     1$ & $  8.22$ & $ -9.99$ & $     1$ & $  8.10$ & $  0.10$ & $     4$ & $ -9.99$ & $ -9.99$ & $     0$\\
&&&&&&&&&&&& \\[-10pt]
\hline
\end{tabular}
\end{table*}

\subsubsection{Magnitudes in the SDSS system}
\label{sec:magnitudes22}

Below, the entries in Tab.~(\ref{tab22}) are described column by column. The entries on the magnitudes and their uncertainties in that table are set to $-9.99$ if the value is not available, and always positive otherwise. Uncertainties, if available, always come from the same source as the value for the magnitude itself.

{\bf Column~1} lists the running number of the galaxy.

{\bf Column~2} lists the apparent $u$-band magnitude, $m_u$.

{\bf Column~3} quotes the uncertainty of $m_u$.

{\bf Column~4} lists the source for the values for $m_u$. The number in this column is set to `0' if no data on $m_u$ is available. A `1' indicates that the value for $m_u$ comes from the tenth data release of the SDSS\footnote{https://www.sdss3.org/dr10/}. We correct this data for redshift and Galactic foreground extinction. A `2' in Column~4 indicates that the galaxy is a member of the galaxy cluster Abell~496 from \citet{Chilingarian2008}, who already corrected this data for extinction and redshift. A number of the form `10xx' indicates that the value for $m_u$ has been calculated as described in Section~(\ref{sec:lumcorrelation}).

{\bf Column~5} lists the apparent $g$-band magnitudes, $m_g$.

{\bf Column~6} quotes the uncertainty of $m_g$.

{\bf Column~7} lists the source for the values for $m_g$. The allocation of numbers to sources is the same as in column~4.

{\bf Column~8} lists the apparent $r$-band magnitudes, $m_r$.

{\bf Column~9} quotes the uncertainty of $m_r$.

{\bf Column~10} lists the sources for the values for $m_r$. The allocation of numbers to sources is the same as in column~4.

{\bf Column~11} lists the apparent $i$-band magnitudes, $m_i$.

{\bf Column~12} quotes the uncertainty of $m_i$.

{\bf Column~13} lists the sources for the values for $m_i$. The allocation of numbers to sources is the same as in column~4.

{\bf Column~14} lists the apparent $z$-band magnitudes, $m_z$.

{\bf Column~15} quotes the uncertainty of $m_z$.

{\bf Column~16} lists the sources for the values for $m_z$. The allocation of numbers to sources is the same as in column~4.

\begin{table*}
\caption{\label{tab22} SDSS-system magnitudes of the galaxies in this catalogue. A portion of the Table is shown here for guidance regarding its contents and form. A detailed description of the contents of this table is given in Section~(\ref{sec:magnitudes2}). The table in its entirety can be downloaded at http://www.astro-udec.cl/mf/catalogue/}
\centering
\vspace{2mm}
\begin{tabular}{rrrrrrrrrrrrrrrrrr}
\hline
&&&&&&&&&&&&&&& \\[-10pt]
id     & $ m_u $  & $dm_u$   & s.   & $ m_g $  & $dm_g  $ & s.   & $ m_r$   & $dm_r  $ & s.   & $  m_i $ & $ dm_i $ & s.   & $ m_z$   & $ dm_z $ & s. \\
        &  $[{\rm mag}]$ & $[{\rm mag}]$   &                &$[{\rm mag}]$&$[{\rm mag}]$      &               &  $[{\rm mag}]$    &  $[{\rm mag}]$    &               &  $[{\rm mag}]$    & $[{\rm mag}]$     &               &$[{\rm mag}]$    &  $[{\rm mag}]$      &             \\
\hline
$ 501$ & $ 15.33$ & $  0.01$ & $     1$ & $ 13.60$ & $  0.00$ & $     1$ & $ 12.85$ & $  0.00$ & $     1$ & $ 12.45$ & $  0.00$ & $     1$ & $ 12.23$ & $  0.00$ & $     1$\\
$ 502$ & $ 14.19$ & $  0.00$ & $     1$ & $ 12.95$ & $  0.00$ & $     1$ & $ 12.32$ & $  0.00$ & $     1$ & $ 13.43$ & $  0.01$ & $     1$ & $ 11.81$ & $  0.00$ & $     1$\\
$ 503$ & $ 16.02$ & $  0.01$ & $     1$ & $ 14.34$ & $  0.00$ & $     1$ & $ 13.57$ & $  0.00$ & $     1$ & $ 13.27$ & $  0.00$ & $     1$ & $ 13.02$ & $  0.00$ & $     1$\\
$ 504$ & $ 13.57$ & $  0.43$ & $  1002$ & $ 11.74$ & $  0.43$ & $  1002$ & $ 11.01$ & $  0.43$ & $  1030$ & $ -9.99$ & $ -9.99$ & $     0$ & $ -9.99$ & $ -9.99$ & $     0$\\
$ 505$ & $ 15.02$ & $  0.00$ & $     1$ & $ 13.67$ & $  0.00$ & $     1$ & $ 13.20$ & $  0.00$ & $     1$ & $ 12.94$ & $  0.00$ & $     1$ & $ 12.76$ & $  0.00$ & $     1$\\
$ 506$ & $ 14.35$ & $  0.00$ & $     1$ & $ 12.71$ & $  0.00$ & $     1$ & $ 11.97$ & $  0.00$ & $     1$ & $ 11.54$ & $  0.00$ & $     1$ & $ 11.29$ & $  0.00$ & $     1$\\
$ 507$ & $ 15.27$ & $  0.00$ & $     1$ & $ 14.33$ & $  0.00$ & $     1$ & $ 13.78$ & $  0.00$ & $     1$ & $ 13.60$ & $  0.00$ & $     1$ & $ 13.40$ & $  0.00$ & $     1$\\
$ 508$ & $ 16.42$ & $  0.01$ & $     1$ & $ 14.80$ & $  0.00$ & $     1$ & $ 14.09$ & $  0.00$ & $     1$ & $ 13.74$ & $  0.00$ & $     1$ & $ 13.44$ & $  0.00$ & $     1$\\
$ 509$ & $ 13.76$ & $  3.05$ & $  1002$ & $ 13.56$ & $  1.70$ & $  1002$ & $ 12.36$ & $  1.25$ & $  1030$ & $ -9.99$ & $ -9.99$ & $     0$ & $ -9.99$ & $ -9.99$ & $     0$\\
$ 510$ & $  7.06$ & $  0.43$ & $  1012$ & $  5.59$ & $  0.43$ & $  1012$ & $  5.59$ & $  0.43$ & $  1030$ & $ -9.99$ & $ -9.99$ & $     0$ & $ -9.99$ & $ -9.99$ & $     0$\\
&&&&&&&&&&&&&&&  \\[-10pt]
\hline
\end{tabular}
\end{table*}

\subsubsection{luminosities}
\label{sec:luminosities}

The structure of Tables~(\ref{tab31}) and~(\ref{tab32}) is the same as the structure of Tabs.~(\ref{tab21}) and~(\ref{tab22}), but with luminosities instead of apparent magnitudes. The entries in the columns indicating the source of the data are the same as the respective entries in Tabs.~(\ref{tab21}) and~(\ref{tab22}), even though this neglects the different reliability of the distance estimates in Tab.~(\ref{tab12}), which are needed for the transition from apparent magnitudes to luminosities. The entries that table are set to $-9.99$ if some value is not available, and always positive otherwise.

\begin{table*}
\caption{\label{tab31} The standard luminosities of the galaxies in this catalogue. A portion of the Table is shown here for guidance regarding its contents and form. The contents follow the same pattern as the entries in tables~(\ref{tab21}) and~(\ref{tab22}), but for the luminosities instead of the apparent magnitudes. A detailed description of the contents of this table is given in Section~(\ref{sec:magnitudes2}). The table in its entirety can be downloaded at http://www.astro-udec.cl/mf/catalogue/}
\centering
\vspace{2mm}
\begin{tabular}{rrrrrrrrrrrrr}
\hline
&&&&&&&&&&&& \\[-10pt]
id     & $ L_U$ & $ dL_U$    & s.   & $ L_B$   & $ dL_B$  & s.   & $ L_V $  & $ dL_V$  & s.   & $ L_I $  & $dL_I$   & source\\
        &$[\lg(L_{\odot})]$&$[\lg(L_{\odot})]$    &              & $[\lg(L_{\odot})]$ &$[\lg(L_{\odot})]$  &               & $[\lg(L_{\odot})]$&$[\lg(L_{\odot})]$  &               &$[\lg(L_{\odot})]$&$[\lg(L_{\odot})]$&           \\
\hline
$ 501$ & $  9.48$ & $  0.20$ & $  1012$ & $  9.56$ & $ -9.99$ & $     1$ & $  9.66$ & $  0.19$ & $  1023$ & $  9.71$ & $  0.18$ & $  1045$\\
$ 502$ & $  9.86$ & $ -9.99$ & $     1$ & $  9.84$ & $ -9.99$ & $     1$ & $  9.83$ & $ -9.99$ & $     1$ & $  9.57$ & $  0.18$ & $  1045$\\
$ 503$ & $  9.55$ & $  0.20$ & $  1012$ & $  9.57$ & $ -9.99$ & $     1$ & $  9.73$ & $  0.19$ & $  1023$ & $  9.74$ & $  0.18$ & $  1045$\\
$ 504$ & $ -9.99$ & $ -9.99$ & $     0$ & $ 10.88$ & $ -9.99$ & $     1$ & $ 10.99$ & $ -9.99$ & $     1$ & $ -9.99$ & $ -9.99$ & $     0$\\
$ 505$ & $  9.15$ & $ -9.99$ & $     1$ & $  9.36$ & $ -9.99$ & $     1$ & $  9.35$ & $ -9.99$ & $     1$ & $  9.43$ & $  0.18$ & $  1045$\\
$ 506$ & $  9.47$ & $ -9.99$ & $     1$ & $  9.73$ & $ -9.99$ & $     1$ & $  9.79$ & $ -9.99$ & $     1$ & $  9.82$ & $  0.18$ & $  1045$\\
$ 507$ & $  9.81$ & $ -9.99$ & $     1$ & $  9.65$ & $ -9.99$ & $     1$ & $  9.61$ & $ -9.99$ & $     1$ & $  9.70$ & $  0.18$ & $  1045$\\
$ 508$ & $  9.38$ & $  0.20$ & $  1012$ & $  9.42$ & $ -9.99$ & $     1$ & $  9.53$ & $  0.19$ & $  1023$ & $  9.57$ & $  0.18$ & $  1045$\\
$ 509$ & $ -9.99$ & $ -9.99$ & $     0$ & $  3.49$ & $ -9.99$ & $     4$ & $  3.67$ & $ -9.99$ & $     4$ & $ -9.99$ & $ -9.99$ & $     0$\\
$ 510$ & $  8.60$ & $ -9.99$ & $     1$ & $  8.68$ & $ -9.99$ & $     1$ & $  8.47$ & $ -9.99$ & $     4$ & $ -9.99$ & $ -9.99$ & $     0$\\
&&&&&&&&&&&& \\[-10pt]
\hline
\end{tabular}
\end{table*}

\begin{landscape}
\begin{table}
\caption{\label{tab32} The SDSS luminosities of the galaxies in this catalogue. A portion of the Table is shown here for guidance regarding its contents and form. The contents follow the same pattern as the entries in Tab.~(\ref{tab21}) and~(\ref{tab22}), but for the luminosities instead of the apparent magnitudes. A detailed description of the contents of this table is given in Section~(\ref{sec:magnitudes2}). The table in its entirety can be downloaded at http://www.astro-udec.cl/mf/catalogue/}
\centering
\vspace{2mm}
\begin{tabular}{rrrrrrrrrrrrrrrrrr}
\hline
&&&&&&&&&&&&&&& \\[-10pt]
id     & $ L_u $  & $dL_u$   & s. & $ L_g $  & $dL_g  $ & s.   & $ L_r$   & $dL_r  $ & s.   & $  L_i $ & $ dL_i $ & s.   & $ L_z$   & $ dL_z $ & s. \\
       &  $[\lg(L_{\odot})]$ &  $[\lg(L_{\odot})]$  &             &  $[\lg(L_{\odot})]$ &  $[\lg(L_{\odot})]$  &              &  $[\lg(L_{\odot})]$ &  $[\lg(L_{\odot})]$ &               &  $[\lg(L_{\odot})]$& $[\lg(L_{\odot})]$ &              &  $[\lg(L_{\odot})]$ &  $[\lg(L_{\odot})]$  &            \\
\hline
$ 501$ & $  9.29$ & $  0.00$ & $     1$ & $  9.48$ & $  0.00$ & $     1$ & $  9.58$ & $  0.00$ & $     1$ & $  9.70$ & $  0.00$ & $     1$ & $  9.78$ & $  0.00$ & $     1$\\
$ 502$ & $  9.71$ & $  0.00$ & $     1$ & $  9.70$ & $  0.00$ & $     1$ & $  9.77$ & $  0.00$ & $     1$ & $  9.28$ & $  0.00$ & $     1$ & $  9.92$ & $  0.00$ & $     1$\\
$ 503$ & $  9.37$ & $  0.00$ & $     1$ & $  9.54$ & $  0.00$ & $     1$ & $  9.65$ & $  0.00$ & $     1$ & $  9.73$ & $  0.00$ & $     1$ & $  9.82$ & $  0.00$ & $     1$\\
$ 504$ & $ 10.57$ & $  0.17$ & $  1002$ & $ 10.79$ & $  0.17$ & $  1002$ & $ 10.89$ & $  0.17$ & $  1030$ & $ -9.99$ & $ -9.99$ & $     0$ & $ -9.99$ & $ -9.99$ & $     0$\\
$ 505$ & $  9.32$ & $  0.00$ & $     1$ & $  9.36$ & $  0.00$ & $     1$ & $  9.36$ & $  0.00$ & $     1$ & $  9.42$ & $  0.00$ & $     1$ & $  9.48$ & $  0.00$ & $     1$\\
$ 506$ & $  9.42$ & $  0.00$ & $     1$ & $  9.58$ & $  0.00$ & $     1$ & $  9.68$ & $  0.00$ & $     1$ & $  9.81$ & $  0.00$ & $     1$ & $  9.90$ & $  0.00$ & $     1$\\
$ 507$ & $  9.77$ & $  0.00$ & $     1$ & $  9.64$ & $  0.00$ & $     1$ & $  9.67$ & $  0.00$ & $     1$ & $  9.69$ & $  0.00$ & $     1$ & $  9.77$ & $  0.00$ & $     1$\\
$ 508$ & $  9.21$ & $  0.00$ & $     1$ & $  9.35$ & $  0.00$ & $     1$ & $  9.44$ & $  0.00$ & $     1$ & $  9.54$ & $  0.00$ & $     1$ & $  9.65$ & $  0.00$ & $     1$\\
$ 509$ & $  3.77$ & $  1.22$ & $  1002$ & $  3.35$ & $  0.68$ & $  1002$ & $  3.64$ & $  0.50$ & $  1030$ & $ -9.99$ & $ -9.99$ & $     0$ & $ -9.99$ & $ -9.99$ & $     0$\\
$ 510$ & $  8.50$ & $  0.17$ & $  1012$ & $  8.58$ & $  0.17$ & $  1012$ & $  8.39$ & $  0.17$ & $  1030$ & $ -9.99$ & $ -9.99$ & $     0$ & $ -9.99$ & $ -9.99$ & $     0$\\
&&&&&&&&&&&&&&&  \\[-10pt]
\hline
\end{tabular}
\end{table}
\end{landscape}

\section[Colours]{Colours}
\label{sec:colours}

\subsection{General procedures}
\label{sec:colours1}

In Tables~(\ref{tab41}) and~(\ref{tab42}), we list selected colours of the galaxies. For the sake of internal consistency within our catalog, we preferentially calculate the colours from the magnitudes in Tables~(\ref{tab21}) and~(\ref{tab22}). These estimates for the colours are therefore based on the same assumptions made for in Tables~(\ref{tab21}) and~(\ref{tab22}) for homogenising the data on the magnitudes and correcting them for extinction and redshift (see Section~\ref{sec:magnitudes}). Note that we only calculate a colour from the magnitudes if the two magnitudes are coming from the same source. A single source for the two magnitudes either indicates that the magnitudes are coming from a homogenised database, or can be taken as an indicator that a single team has observed the given galaxy with a single telescope and has reduced the data homogeneously. The uncertainties to these estimates for the colours are calculated via gaussian error propagation, if uncertainties for $m_X$ and $m_Y$ are available. If the resulting uncertainty is larger than 0.6 mag, the $(X-Y)$-colour calculated by subtracting $m_X$ from $m_Y$ is not entered in Tables~(\ref{tab41}) or~(\ref{tab42}), since it is of no use due to its uncertainty.

\subsection{Detailed description of the data}
\label{sec:colours2}

\subsubsection{Colours in the standard system}
\label{sec:colours21}

Below, we explain the entries in Tab.~(\ref{tab41}) column by column. The galaxies are ordered by ascending right ascension $\alpha(2000.0)$ The entries that table are set to $-9.99$ if some value is not available. The values are corrected for extinction and redshift, and their uncertainties, if available, come from the same source as the value itself.

{\bf Column~1} lists the running number of the galaxy.

{\bf Column~2} lists the $(U-B)$-colour of the galaxy. 

{\bf Column~3} quotes the uncertainty of $(U-B)$.

{\bf Column~4} lists the sources for $(U-B)$. The number in this column is set to `0' if no data on $(U-B)$ is available. A `1' indicates that the source for $(U-B)$ is HyperLeda. This is the primary source on $(U-B)$, since HyperLeda provides for galaxies of all luminosities a wealth of data on $(U-B)$ that have already been homogenised and corrected for extinction and redshift. A `2' indicates that $(U-B)$ has been calculated based on data on $m_U$ and $m_B$ from \citet{Eisenhardt2007} for the Coma Cluster.

{\bf Column~5} lists the $(B-V)$-colour of the galaxy. 

{\bf Column~6} quotes the uncertainty of $(B-V)$.

{\bf Column~7} lists the sources for $(B-V)$. The number in this column is set to `0' if no data on $(B-V)$ is available. A `1' indicates that the source for $(B-V)$ is HyperLeda. For the same reasons as for $(U-B)$, HyperLeda is the primary source on $(B-V)$. A `2' indicates that the colour has been calculated based on data on $m_B$ and $m_V$ from \citet{Eisenhardt2007}. A `3' indicates that the colour has been calculated based on data on $m_B$ and $m_V$ from \citet{Karick2003} for the Fornax Cluster. A `4' indicates that $(B-V)$ comes from \citet{Mateo1998}, who collected properties of galaxies in the Local Group.

{\bf Column~8} lists the $(V-I)$-colour of the galaxy. 

{\bf Column~9} quotes the uncertainty of $(V-I)$.

{\bf Column~10} lists the sources for $(V-I)$. The number in this column is set to `0' if no data on $(V-I)$ is available. A `2' indicates that the colour has been calculated based on data on $m_V$ and $m_I$ from \citet{Eisenhardt2007} for the Coma Cluster. A `3' indicates that the colour has been calculated based on data from \citet{Karick2003}. A `4' indicates that the colour has been calculated based on data on $m_V$ and $m_I$ from \citet{Mieske2007} for the Fornax Cluster. As with the magnitudes, we prefer data on $(V-I)$ from \citet{Karick2003} over those from \citet{Mieske2007}. A `5' indicates that the colour has been calculated based data on $m_V$ and $m_I$ from \citet{Lieder2012} for the Virgo Cluster. A `6' indicates that the colour has been calculated based on data on $m_V$ and $m_I$ from \citet{Misgeld2008} for the Hercules Cluster. A `7' indicates that the colour has been calculated based data on $m_V$ and $m_I$ from \citet{Misgeld2009} for the Centaurus Cluster.

\begin{table*}
\caption{\label{tab41} Selected standard-system colours of the galaxies in this catalogue. A portion of the Table is shown here for guidance regarding its contents and form. A detailed description of the contents of this table is given in Section~(\ref{sec:colours2}). The table in its entirety can be downloaded at http://www.astro-udec.cl/mf/catalogue/}
\centering
\vspace{2mm}
\begin{tabular}{rrrrrrrrrr}
\hline
&&&&&&&&& \\[-10pt]
id     & $ (U-B)$ & $ d(U-B)$& s.   & $ (B-V)$ & $ d(B-V)$& s.   & $ (V-I)$ & $ d(V-I)$& s. \\
       &  $[{\rm mag}]$       & $[{\rm mag}]$       &              &  $[{\rm mag}]$       &$[{\rm mag}]$&               &  $[{\rm mag}]$   &  $[{\rm mag}]$     &            \\
\hline
$ 501$ & $ -9.99$ & $ -9.99$ & $     0$ & $ -9.99$ & $ -9.99$ & $     0$ & $ -9.99$ & $ -9.99$ & $     0$\\
$ 502$ & $  0.09$ & $ -9.99$ & $     1$ & $  0.63$ & $ -9.99$ & $     1$ & $ -9.99$ & $ -9.99$ & $     0$\\
$ 503$ & $ -9.99$ & $ -9.99$ & $     0$ & $ -9.99$ & $ -9.99$ & $     0$ & $ -9.99$ & $ -9.99$ & $     0$\\
$ 504$ & $ -9.99$ & $ -9.99$ & $     0$ & $  0.92$ & $ -9.99$ & $     1$ & $ -9.99$ & $ -9.99$ & $     0$\\
$ 505$ & $  0.64$ & $ -9.99$ & $     1$ & $  0.64$ & $ -9.99$ & $     1$ & $ -9.99$ & $ -9.99$ & $     0$\\
$ 506$ & $  0.78$ & $ -9.99$ & $     1$ & $  0.78$ & $ -9.99$ & $     1$ & $ -9.99$ & $ -9.99$ & $     0$\\
$ 507$ & $ -0.26$ & $ -9.99$ & $     1$ & $  0.53$ & $ -9.99$ & $     1$ & $ -9.99$ & $ -9.99$ & $     0$\\
$ 508$ & $ -9.99$ & $ -9.99$ & $     0$ & $ -9.99$ & $ -9.99$ & $     0$ & $ -9.99$ & $ -9.99$ & $     0$\\
$ 509$ & $ -9.99$ & $ -9.99$ & $     0$ & $ -9.99$ & $ -9.99$ & $     0$ & $ -9.99$ & $ -9.99$ & $     0$\\
$ 510$ & $  0.34$ & $ -9.99$ & $     1$ & $  0.99$ & $  0.05$ & $     5$ & $ -9.99$ & $ -9.99$ & $     0$\\
&&&&&&&&& \\[-10pt]
\hline
\end{tabular}
\end{table*}

\subsubsection{Colours in the SDSS system}
\label{sec:colours22}

Below, we explain the entries in Tab.~(\ref{tab42}) column by column. The galaxies are ordered by ascending right ascension $\alpha(2000.0)$ The entries that table are set to $-9.99$ if some value is not available. The values are corrected for extinction and redshift, and their uncertainties, if available, come from the same source as the value itself.

{\bf Column~1} lists the running number of the galaxy.

{\bf Column~2} lists the $(u-g)$-colour of the galaxy. 

{\bf Column~3} quotes the uncertainty of $(u-g)$.

{\bf Colomn~4} lists the sources for $(u-g)$. The number in this column is set to `0' if no data on $(u-g)$ is available. A `1' indicates that the colour has been calculated based on data on $m_u$ and $m_g$ from the SDSS. A `2' indicates that the colour has been calculated from data on $m_u$ and $m_g$ from \citet{Chilingarian2008} on the galaxy cluster Abell~496.

{\bf Column~5} lists the $(g-r)$-colour of the galaxy. 

{\bf Column~6} quotes the uncertainty of $(g-r)$.

{\bf Colomn~7} lists the sources for $(g-r)$. The allocation of numbers to sources is the same as in column~4.

{\bf Column~8} lists the $(g-i)$-colour of the galaxy. 

{\bf Column~9} quotes the uncertainty of $(g-i)$.

{\bf Colomn~10} lists the sources for $(g-i)$. The allocation of numbers to sources is the same as in column~4.

{\bf Column~11} lists the $(g-z)$-colour of the galaxy. 

{\bf Column~12} quotes the uncertainty of $(g-z)$.

{\bf Colomn~13} lists the sources for $(g-z)$. The allocation of numbers to sources is the same as in column~4.

\begin{table*}
\caption{\label{tab42} Selected SDSS colours of the galaxies in this catalogue. A portion of the Table is shown here for guidance regarding its contents and form. A detailed description of the contents of this table is given in Section~(\ref{sec:colours2}). The table in its entirety can be downloaded at http://www.astro-udec.cl/mf/catalogue/.}
\centering
\vspace{2mm}
\begin{tabular}{rrrrrrrrrrrrr}
\hline
&&&&&&&&&&&& \\[-10pt]
id     & $ (u-g)$ & $d(u-g)$ & s.   & $ (g-r)$ & $d(g-r)$ & s.   & $ (g-i)$ & $d(g-i)$ & s.   & $ (g-z)$ & $d(g-z)$ & s. \\
       &  $[{\rm mag}]$     &$[{\rm mag}]$       &              &  $[{\rm mag}]$&  $[{\rm mag}]$     &               &  $[{\rm mag}]$&  $[{\rm mag}]$&               &  $[{\rm mag}]$     &  $[{\rm mag}]$       &            \\
\hline
$ 501$ & $  1.73$ & $  0.01$ & $     1$ & $  0.74$ & $  0.00$ & $     1$ & $  1.14$ & $  0.00$ & $     1$ & $  1.36$ & $  0.00$ & $     1$\\
$ 502$ & $  1.24$ & $  0.00$ & $     1$ & $  0.63$ & $  0.00$ & $     1$ & $ -0.47$ & $  0.01$ & $     1$ & $  1.15$ & $  0.00$ & $     1$\\
$ 503$ & $  1.68$ & $  0.01$ & $     1$ & $  0.76$ & $  0.00$ & $     1$ & $  1.07$ & $  0.00$ & $     1$ & $  1.32$ & $  0.00$ & $     1$\\
$ 504$ & $ -9.99$ & $ -9.99$ & $     0$ & $ -9.99$ & $ -9.99$ & $     0$ & $ -9.99$ & $ -9.99$ & $     0$ & $ -9.99$ & $ -9.99$ & $     0$\\
$ 505$ & $  1.36$ & $  0.00$ & $     1$ & $  0.47$ & $  0.00$ & $     1$ & $  0.73$ & $  0.00$ & $     1$ & $  0.91$ & $  0.00$ & $     1$\\
$ 506$ & $  1.64$ & $  0.00$ & $     1$ & $  0.73$ & $  0.00$ & $     1$ & $  1.17$ & $  0.00$ & $     1$ & $  1.41$ & $  0.00$ & $     1$\\
$ 507$ & $  0.94$ & $  0.00$ & $     1$ & $  0.55$ & $  0.00$ & $     1$ & $  0.73$ & $  0.00$ & $     1$ & $  0.93$ & $  0.00$ & $     1$\\
$ 508$ & $  1.62$ & $  0.01$ & $     1$ & $  0.72$ & $  0.00$ & $     1$ & $  1.06$ & $  0.00$ & $     1$ & $  1.36$ & $  0.00$ & $     1$\\
$ 509$ & $ -9.99$ & $ -9.99$ & $     0$ & $ -9.99$ & $ -9.99$ & $     0$ & $ -9.99$ & $ -9.99$ & $     0$ & $ -9.99$ & $ -9.99$ & $     0$\\
$ 510$ & $ -9.99$ & $ -9.99$ & $     0$ & $ -9.99$ & $ -9.99$ & $     0$ & $ -9.99$ & $ -9.99$ & $     0$ & $ -9.99$ & $ -9.99$ & $     0$\\
&&&&&&&&&&&& \\[-10pt]
\hline
\end{tabular}
\end{table*}

\section[Structural properties]{Structural properties and internal dynamics}
\label{sec:dynamics}

\subsection{General procedures}
\label{sec:dynamics1}

In Tab.~(\ref{tab51}), we list structural properties, namely the effective half-light radii, internal velocity dispersions and the maximal rotation velocities of the galaxies.

\subsubsection{Half-light radii}
\label{sec:radii}

The isophotes of ETGs are usually well quantified by ellipses that are distinctively different from a perfect circle. This deviation from rotational symmetry can be quantified with the ellipticity $\epsilon$, which is defined here as the ratio between the semi-minor and the semi-major axes of the ellipse that includes half of the total luminosity of the galaxy. A parameter that characterises the extension of the galaxy and does not explicitly depend on $\epsilon$ is given by
\begin{equation}
R_{\rm e}=R_{\rm maj} \, \sqrt{1-\epsilon},
\label{eq:equradius1}
\end{equation}
where $R_{\rm maj}$ is the semi-major axis.
If necessary, Eq.~(\ref{eq:equradius1}) is used to calculate $R_{\rm e}$ from the data found in the literature.

Data on $R_{\rm e}$ given in arc seconds or arc minutes in the literature are converted into parsecs, using the distances listed in Tab.~(\ref{tab12}). If the original source already gives the physical radius for a galaxy, we first convert it into an angular diameter using the distance estimate in the original paper, and then convert it back into an estimate for its physical radius using the distance listed in Tab.~(\ref{tab12}). This is only a minor correction for most of the considered galaxies, but increases the internal consistency of our catalogue.

It should be noted that $R_{\rm e}$ have been estimated from various passbands, ranging from the visual to the near infrared. This is a source for a certain inhomogeneity among the data for $R_{\rm e}$. However, \citet{Ferrarese2006} present data for $R_{\rm e}$ estimated from observations in the $g$-band and the $z$-band, and the the difference between the two estimates is in most cases not very dramatic.

\subsubsection{S\'{e}rsic-profiles}

The differences in the luminosity profiles of ETGs can conveniently be quantified by the S\'{e}rsic-index $n$, i.e. the number $n$ that best describes the overall luminosity profile of the galaxy in a generalised luminosity profile of the form
\begin{equation}
I(R)=I_0 \exp \left[-b_{n}\left(\frac{r}{R_{\rm e}}\right)^{\frac{1}{n}}\right],
\end{equation}
where $I(R)$ is the surface brightness of the galaxy as a function of the radius $R$, $I_0$ is the central surface brightness, $R_{\rm e}$ is the effective radius and $b_{n} \approx 2n-0.324$ \citep{Ciotti1991}.

Observations have shown that $n$ correlates with $R_{\rm e}$. \citet{Caon1993} quantify this relation as
\begin{equation}
\log_{10} (n) = 0.28+0.52 \log_{10}\left(\frac{R_{\rm e}}{{\rm kpc}}\right),
\label{eq:sersic}
\end{equation}
and the scatter $\Delta n$ around this relation is 0.18. We use Eq.~(\ref{eq:sersic}) to calculate $n$, and take $\Delta n=0.18$ as the uncertainty, if we do not find an estimate for $n$ in the literature.

\subsubsection{Internal velocity dispersions}
\label{sec:sigma}

There are different quantities that are widely used for characterising the internal velocity dispersions of galaxies. Among them are the velocity dispersion at the centre of the galaxy, $\sigma_{0}$, and the velocity dispersion within the projected half-light radius, $\sigma_{\rm e}$. For the galaxies in this catalogue, $\sigma_{\rm e}$ is the more widely available quantity, but estimators of the dynamical mass of ETGs are often formulated in terms of $\sigma_{0}$ rather than $\sigma_{\rm e}$. We therefore present both quantities in our catalogue. If $\sigma_{\rm e}$ is available for a given ETG in the catalogue, but $\sigma_{0}$ is not, the relation
\begin{equation}
\log_{10}\left(\frac{\sigma_{0}}{{\rm km/s}}\right)=1.0478 \, \log_{10}\left(\frac{\sigma_{\rm e}}{{\rm km/s}}\right)-0.0909
\label{eq:sigmae}
\end{equation}
is used in order to estimate $\sigma_{0}$ from $\sigma_{\rm e}$. This equation is obtained from a least-squares fit to data on 260 ETGs in \citet{Cappellari2013}, while taking their data on the velocity dispersions within the central parsec of the ETGs as a measure for $\sigma_0$. Swapping $\sigma_{0}$ and $\sigma_{\rm e}$ in the fit leads to
\begin{equation}
\log_{10}\left(\frac{\sigma_{\rm e}}{{\rm km/s}}\right)=0.9251 \, \log_{10}\left(\frac{\sigma_{0}}{{\rm km/s}}\right)+0.1485,
\label{eq:sigma0}
\end{equation}
which is used to estimate $\sigma_{\rm e}$ from $\sigma_0$, if only $\sigma_{0}$ has been found in the literature for a given galaxy.

By construction, Eqs.~({\ref{eq:sigmae}}) and~({\ref{eq:sigma0}}) cannot be used on ETGs with $R_{\rm e}<1$ kpc. We therefore set $\sigma_{0}=\sigma_{\rm e}$ for such ETGs, which is well motivated by the findings that $R_{\rm e}=1$ kpc corresponds to $n \approx 1.9$ according to Eq.~(\ref{eq:sersic}), and that the velocity dispersion profiles of galaxies with $n\apprle 2$ are almost flat within their $R_{\rm e}$ \citep{Graham1997,Simonneau2004}.

\subsubsection{Rotational velocities}
\label{sec:rotation}

Even though random motion dominates over ordered motion in ETGs, many of them still rotate substantially. We therefore provide data on $v_{\rm rot}$ of the galaxies, i.e. the rotational velocities of the galaxies within their $R_{\rm e}$. This data in particular should be taken with some caution, since $v_{\rm rot}$ is not quantified in a uniform way in the literature that was used for this catalogue. We aim at finding an average rotational velocity of the ETG within $R_{\rm e}$.

We come close to this aim with the data on the average ratios between the internal velocity dispersion and rotational velocities of the galaxies in the ATLAS$^{3D}$-sample, which are provided by \citet{Emsellem2011}. We multiply the ratios from their paper with $\sigma_{\rm e}$ in order to estimate $v_{\rm rot}$. The resulting estimates for $v_{\rm rot}$ are quite consistent with the data on $v_{\rm rot}$ of ETGs provided by \citet{Scodeggio1998b}, in the sense that the data from both subsamples occupy the same region in $L_V$-$v_{\rm rot}$ parameter space. At lower luminosities, it is more common to provide (rough) estimates for the maximum rotational velocities, as done \citet{Geha2003} and \citet{McConnachie2012}. Finally, \citet{Toloba2014} provide data on the rotational velocities at the radius $R_{\rm e}$ of the ETGs in their sample.

Despite the noted inhomogeneities in the data on $v_{\rm rot}$, we expect them to give a rather consistent picture of the amount of rotation that can be expected in ETGs. Following an argument in \citet{Geha2003}, the rotational velocity of an ETG is expected to reach its maximum at a radius $0.5 \, R_{\rm e} \apprle R \apprle R_{\rm e}$. This implies that the part of the ETG, where the maximum rotational velocity is reached, has a strong impact on the estimate of $v_{\rm rot}$, if $v_{\rm rot}$ is given by the average rotational velocity within $R_{\rm e}$, as it is done here for the most luminous ETGs. Thus, for the luminous ETGs in our catalogue, $v_{\rm rot}$ is by construction always lower than the maximum rotational velocity in the ETG. However, this is also often the case, if the quoted values for $v_{\rm rot}$ was intended to be an approximation to the maximum rotational velocity in the original literature. For instance, in three out of the four ETGs, in which \citet{Geha2003} actually detected substantial rotation, the values for the maximum rotational velocities are only given as lower limits to the true value, since the rotation curve of those galaxies probably peaks slightly outside the radius to which \citet{Geha2003} had spectroscopic data. Also \citet{McConnachie2012} strives at finding the maximum rotational velocities for the galaxies in his catalogue, but notes that some galaxies may not have been measured out to radii, where the peak of the rotation curve is reached. He also does not correct the data on $v_{\rm rot}$ for inclination, which also implies that the quoted values underestimate the true $v_{\rm rot}$.

\subsubsection{Dynamical masses}

If $\sigma_{\rm 0}$ and $R_{\rm e}$ are known for a ETG in this catalog, its dynamical mass is calculated with
\begin{equation}
M_{\rm dyn}=\frac{K_{\rm V}}{G}\, R_{\rm e}\sigma_{0}^2,
\label{eq:Mdyn}
\end{equation}
where $K_{\rm V}$ is a factor that depends on the shape of the density profile of the ETG, and $G$ is the gravitational constant. We approximate $K_{\rm V}$ with equation~(11) in \citet{Bertin2002}, i.e.
\begin{equation}
K_{\rm V}(n) = \frac{73.32}{10.465+(n-0.94)^2}+0.954,
\end{equation}
where $n$ is the S\'{e}rsic index.

\subsection{Detailed description of the data}
\label{sec:dynamics2}

Below, we explain the entries in Tab.~(\ref{tab51}) column by column. The galaxies are ordered by ascending right ascension $\alpha(2000.0)$.

{\bf Column~1} lists the running number of the galaxy.

{\bf Column~2} lists the physical effective half-light radii, $R_{\rm e}$ of the galaxies. There are only three galaxies in this catalogue without an estimate for $R_{\rm e}$. The value for $R_{\rm e}$ is set to $-9.9$ for them. For the galaxies for which also $L_V$ is known, Fig.~(\ref{fig:L-re}) shows the $R_{\rm e}$ of the ETGs in this catalogue over their $L_V$.

{\bf Column~3} lists the uncertainty of $R_{\rm e}$. If the uncertainty is unknown, it is set to $-9.9$.

{\bf Column~4} lists the sources for $R_{\rm e}$ and its uncertainty. The number in this column is set to `0' if no data on $R_{\rm e}$ is available. A `1' indicates the galaxy is included in the ATLAS$^{\rm 3D}$ survey, and that $R_{\rm e}$ has been taken from \citet{Cappellari2013}. A `2' indicates that $R_{\rm e}$ is based on the data in \citet{Bender1992}. Note that the $R_{\rm e}$ listed in \citet{Bender1992} are considerably larger than the ones listed here, since the distance estimates that they adopt for the galaxies are larger than the ones listed in Tab~(\ref{tab12}). A `3' indicates that $R_{\rm e}$ is based on the angular $R_{\rm e}$ in the $g$-band in \citet{Ferrarese2006}. A `4' indicates that $R_{\rm e}$ is based on the angular $R_{\rm e}$ in \citet{Toloba2014}. A `5' indicates that $R_{\rm e}$ based on the angular $R_{\rm e}$ in \citet{Geha2003}. All galaxies designated with numbers 3, 4 or 5 in this column are part of the Virgo cluster. A `6' indicates that the galaxy is part of the Coma cluster and its $R_{\rm e}$ is based on the physical $R_{\rm e}$ published by \citet{Kourkchi2012b}. A `7' indicates galaxies in the cluster Abell~496 whose $R_{\rm e}$ are based on their angular $R_{\rm e}$ in \citet{Chilingarian2008}. A `8' indicates that $R_{\rm e}$ is based on the angular $R_{\rm e}$ of galaxies published in \citet{Scodeggio1998a}. The values for $R_{\rm e}$ coming from \citet{Scodeggio1998a} tend to be larger than the values for $R_{\rm e}$ from other sources, and also show a larger scatter. This is illustrated in Fig.~(\ref{fig:Scodeggio}). This is the typical behaviour of a sample that is not fully spatially resolved, but is treated as if it was fully resolved. It is therefore best to exclude these galaxies in applications, where knowing an accurate estimate for $R_{\rm e}$ is essential. However, Fig.~(\ref{fig:L-re}) illustrates that the sample of galaxies in this catalog as a whole still defines the known luminosity-radius sequence of ETGs well (cf. for instance figure~1 in \citealt{Misgeld2011}). A `9' indicates that the galaxy is part of the Virgo cluster and its $R_{\rm e}$ is based on the angular $R_{\rm e}$ from \citet{Ferguson1989}. A `10' indicates that $R_{\rm e}$ is estimated from angular semi-major axes published by \citet{McConnachie2012} on galaxies in the Local Group. The semi-major axes are transformed into angular $R_{\rm e}$ with Eq.~(\ref{eq:equradius1}). For this, the information on $\epsilon$ in \citet{McConnachie2012} is used, if \citet{McConnachie2012} provides any. Otherwise, we assume the galaxy to be circular, i.e. $\epsilon=0$. A `11' indicates that the galaxy is part of the Virgo cluster and its $R_{\rm e}$ is based on the physical $R_{\rm e}$ published by \citet{Lieder2012}. A `12' indicates galaxies in the Hydra cluster whose $R_{\rm e}$ is based on the angular $R_{\rm e}$ in \citet{Misgeld2008}. A `13' indicates galaxies in the Centaurus cluster whose $R_{\rm e}$ is based on their angular $R_{\rm e}$ in \citet{Misgeld2009}. A `14' indicates galaxies whose value for $R_{\rm e}$ is based on their angular $R_{\rm e}$ in\citet{Koleva2011}. A `15' indicates that $R_{\rm e}$ is based on the values for the physical $R_{\rm e}$ published by \citet{Guerou2015}.

\begin{figure}
\centering
\includegraphics[scale=0.78]{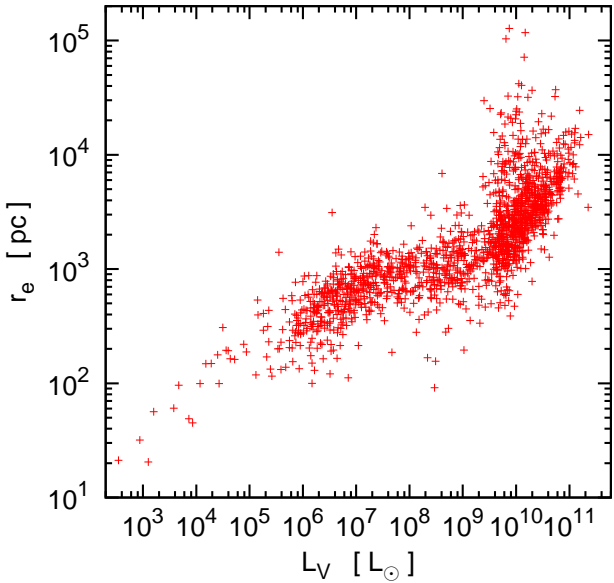}
\caption{\label{fig:L-re} The effective half-light radius, $R_{\rm e}$, over $L_V$ for the ETGs in this catalogue for which both quantities are known.}
\end{figure}

\begin{figure*}
\centering
\includegraphics[scale=0.78]{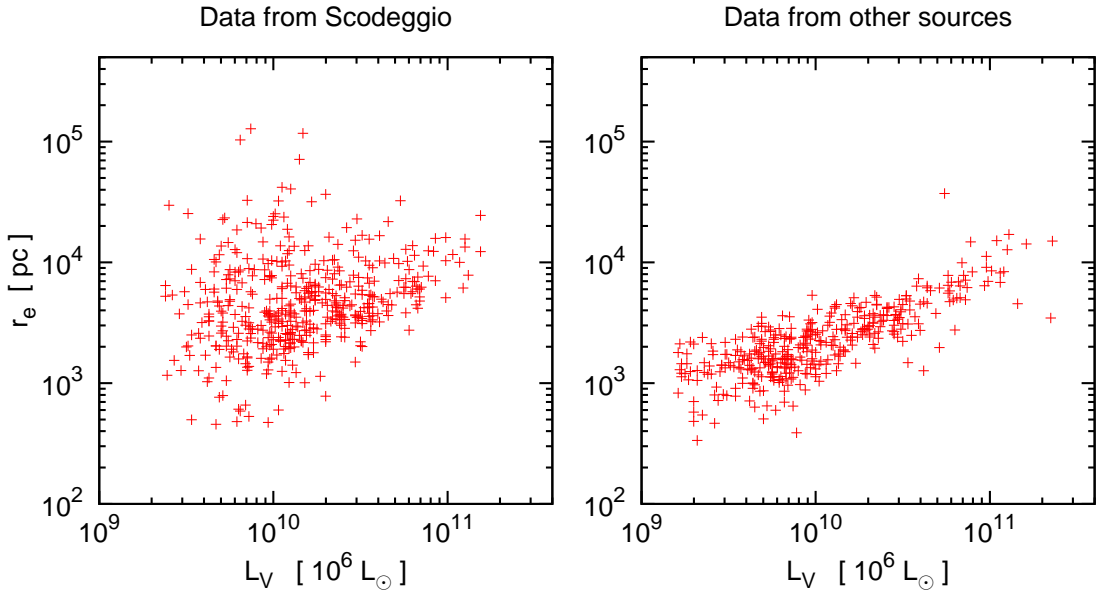}
\caption{\label{fig:Scodeggio} Comparison between values for $R_{\rm e}$ taken from \citet{Scodeggio1998a} and values for $R_{\rm e}$ from other sources. The left panel shows the data on $R_{\rm e}$ from \citet{Scodeggio1998a} The right panel shows all other data in this catalogue on the values $R_{\rm e}$ for ETGs in same luminosity range as the ETGs in\citet{Scodeggio1998a}.}
\end{figure*}

{\bf Column~5} shows the best-fitting S\'{e}rsic-index of the galaxy, $n$. If $n$ is not available, its value is set to $-9.9$. Otherwise, $n>0$.

{\bf Column~6} lists the uncertainty of the S\'{e}rsic-index, if available. Otherwise, its value is set to $-9.9$.

{\bf Column~7} lists the source of the listed values for $n$ and its uncertainty. The number in this column is set to `0' if no data on $n$ is available. A `1' indicates that the source for $n$ is \citet{Krajnovic2013}, a `2' that $n$ is from \citet{Toloba2014}, a `3' that $n$ is from \citet{Geha2003}, a `4' that $n$ from \citet{Kourkchi2012b}, and a `5' that $n$ is from \citet{Guerou2015}. A `101' indicates that $n$ is estimated from $R_{\rm e}$, using Eq.~(\ref{eq:sersic}).

{\bf Column~8} shows $\sigma_{\rm e}$, the internal velocity dispersion within $R_{\rm e}$ of the galaxies. The entry in that column is set to $-9.9$ if $\sigma_{\rm e}$ is unknown.

{\bf Column~9} lists the positive uncertainties of $\sigma_{\rm e}$. The entry in that column is set to $-9.9$, if this uncertainty is unknown.

{\bf Column~10} lists the absolute values of the negative uncertainties of $\sigma_{\rm e}$, since the uncertainties of $\sigma_{\rm e}$ are asymmetric for some galaxies. The entry in that column is set to $-9.9$, if this uncertainty is unknown.

{\bf Column~11} shows the sources for $\sigma_{\rm e}$ and its uncertainties. The number in this column is set to `0' if no data on $\sigma_{\rm e}$ is available. A `1' in column~11 indicates that the galaxy is included in the ATLAS$^{\rm 3D}$ survey, and that the value for $\sigma_{\rm e}$ is taken from \citet{Cappellari2013}. A `2' indicates that $\sigma_{\rm e}$ is from \citet{Forbes2011}, a `3'  that $\sigma_{\rm e}$ is from \citet{Guerou2015}, a `4' that $\sigma_{\rm e}$ is from \citet{Toloba2014}, a `5' that $\sigma_{\rm e}$ is from \citet{Geha2003}, a `6' that $\sigma_{\rm e}$ is from \citet{Kourkchi2012b}, a `7' that $\sigma_{\rm e}$ is from \citet{Chilingarian2008}, a `8' that $\sigma_{\rm e}$ is from \citet{Scodeggio1998b}, a `9' that $\sigma_{\rm e}$ is from \citet{Thomas2005}, a `10' that $\sigma_{\rm e}$ is from \citet{Matkovic2009}, and a `11' that $\sigma_{\rm e}$ is from \citet{SanchezBlazquez2006}. A `12' indicates that the galaxy is part of the Local Group and its $\sigma_{\rm e}$ is based on the internal velocity dispersion given in \citet{McConnachie2012}, which are not strictly values for $\sigma_{\rm e}$. However, since the galaxies from \citet{McConnachie2012} are usually small galaxies with small $R_{\rm e}$ and therefore low $n$, the velocity dispersions quoted for them should nevertheless be good representations of their true $\sigma_{\rm e}$ within the uncertainties due to the difficult observations (see section \ref{sec:sigma}). A`13' indicates that the galaxy is part of the Local group and that its value for $\sigma_{\rm e}$ is taken from HyperLeda.  A `14' indicates that $\sigma_{\rm e}$ is from \citet{Koleva2011}. A `101' indicates that no value for $\sigma_{\rm e}$ has been published, while a value for $\sigma_0$ is available. It is then assumed that $\sigma_{\rm e}=\sigma_0$, if $R_{\rm e}\le1$ kpc. If $R_{\rm e}>1$ kpc, $\sigma_{\rm e}$ is calculated from $\sigma_0$ with Eq.~(\ref{eq:sigmae}).

{\bf Column~12} shows $\sigma_0$, the internal velocity dispersion at the centre of the galaxies. The entry in that column is set to $-9.9$ if $\sigma_0$ is unknown.

{\bf Column~13} lists the positive uncertainties of $\sigma_0$. The entry in that column is set to $-9.9$, if this uncertainty is unknown.

{\bf Column~14} lists the absolute values of the negative uncertainties of $\sigma_0$, since the uncertainties of $\sigma_0$ are asymmetric for some galaxies. The entry in that column is set to $-9.9$, if this uncertainty is unknown.

{\bf Column~15} shows the sources for $\sigma_0$ and its uncertainties. The number in this column is set to `0' if no data on $\sigma_0$ is available. A `1' indicates that the galaxy is included in the ATLAS$^{\rm 3D}$ survey, and that the value for $\sigma_0$ is taken from \citet{Cappellari2013}. A `2' indicates that the value for $\sigma_0$ is taken from \citet{Bender1992}. A `101' indicates that no value for $\sigma_0$ has been published, while a value for $\sigma_{\rm e}$ is available. It is then assumed that $\sigma_0=\sigma_{\rm e}$, if $R_{\rm e}\le1$ kpc. If $R_{\rm e}>1$ kpc, $\sigma_0$ is calculated from $\sigma_{\rm e}$ with Eq.~(\ref{eq:sigma0}).
 
{\bf Column~16} shows $v_{\rm rot}$, the rotational velocities of the galaxies. If there has been an attempt to find a signature of rotation in a given ETG, but none was found within the measurement uncertainty, the entry in that column is set to $v_{\rm rot}=0.0$ km/s. If no information on $v_{\rm rot}$ is available, the entry in column~16 is set to $-9.9$.

{\bf Column~17} lists the positive uncertainties of $v_{\rm rot}$. The entry in that column is set to $-9.9$ if this uncertainty is unknown.

{\bf Column~18} lists the absolute values of the negative uncertainties of $v_{\rm rot}$. The entry in that column is set to $-9.9$ if this uncertainty is unknown.

{\bf Column~19} shows the sources for $v_{\rm rot}$ and its uncertainties. The data on $v_{\rm rot}$ is not very homogeneous, as detailed in Section~(\ref{sec:rotation}). The number in this column is set to `0' if no data on $v_{\rm rot}$ is available. A `1' in column~19 indicates that the galaxy is included in the ATLAS$^{\rm 3D}$ survey and the estimate of $v_{\rm rot}$ is based on the data given in \citet{Emsellem2011}. A `2' indicates that the estimate for $v_{\rm rot}$ is taken from \citet{Toloba2014}. A `3' indicates that the estimate for $v_{\rm rot}$ is taken from \citet{Geha2003}. A `4' indicates that $v_{\rm rot}$ is based on the values for the rotational velocities from \citet{McConnachie2012}. These estimates for $v_{\rm rot}$ are already by themselves not very homogeneous, as is detailed in \citet{McConnachie2012}. A `5' in column~19 indicates that $v_{\rm rot}$ is from HyperLeda. The quoted value is the maximum rotational velocity. A `6' indicates that $v_{\rm rot}$ is from \citet{Scodeggio1998b}. 

{\bf Column~20} lists the estimates for the dynamical mass of the ETGs, $M_{\rm dyn}$. The number actually shown is the logarithm to the base of 10 of $M_{\rm dyn}$ in Solar units. If $M_{\rm dyn}$ is unknown, the number is set to $-99.99$. For the galaxies for which also $L_V$ is known, Fig.~(\ref{fig:L-MLdyn}) shows the $M_{\rm dyn}/L_V$-ratio of the ETGs in this catalogue over their $L_V$. The high $M_{\rm dyn}/L_V$-ratios of ETGs with low luminosities shown in this figure are well established. They are interpreted as non-equilibrium dynamics due to harassment through tidal fields (e.g. \citealt{Kroupa1997}), the presence of a large amount of non-baryonic dark matter (e.g \citealt{Wolf2010, Tollerud2012}), or as a modification of the Newtonian law of gravity at very low accelerations (e.g. \citealt{Angus2008,McGaugh2013}).

{\bf Column~21} lists the logarithmic uncertainty to the values on $M_{\rm dyn}$, if available. If the uncertainty of $M_{\rm dyn}$ is unknown, the value in column~21 is set to $-99.99$. Otherwise, the value in column~21 is always positive.

{\bf Column~22} lists the sources for the estimates on $M_{\rm dyn}$ and their uncertainties. The number in this column is set to `0' if no data on $M_{\rm dyn}$ is available.  A `1' indicates that the galaxy is included in the ATLAS$^{\rm 3D}$ survey, and that the value for $M_{\rm dyn}$ is taken from \citet{Cappellari2013}. A `101' indicates that $M_{\rm dyn}$ is calculated with Eq.~(\ref{eq:Mdyn}) from the data on $R_{\rm e}$, $n$ and $\sigma_{0}$ in Tab.~(\ref{tab51}).

\begin{figure}
\centering
\includegraphics[scale=0.78]{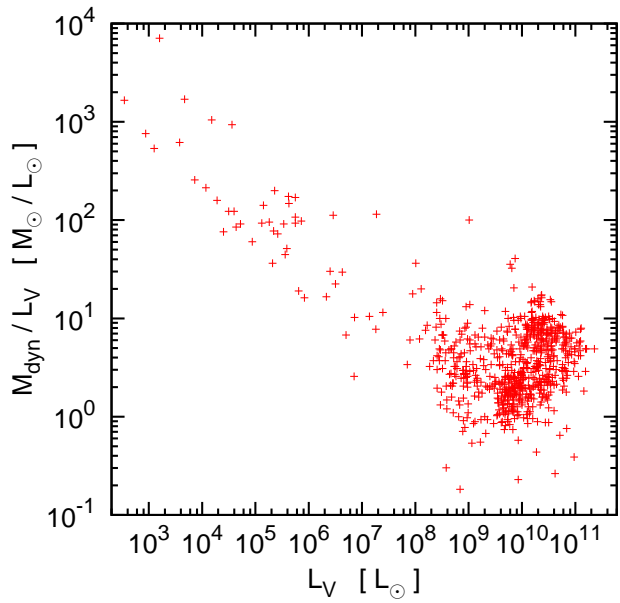}
\caption{\label{fig:L-MLdyn} The ratio between the mass estimated from the internal dynamics of the galaxies and their luminosity, $M_{\rm dyn}/L_V$, over $L_V$ for the ETGs in this catalogue for which both quantities are known.}
\end{figure}

\begin{landscape}
\begin{table}
\caption{\label{tab51} Structural properties and internal dynamics of the galaxies in this catalogue. A portion of the Table is shown here for guidance regarding its contents and form. A detailed description of the contents of this table is given in Section~(\ref{sec:dynamics2}). The table in its entirety can be downloaded at http://www.astro-udec.cl/mf/catalogue/}
\centering
\vspace{2mm}
\begin{tabular}{rrrrrrrrrrrrrrrrrrrrrrrr}
\hline
&&&&&&&&&&&&&&&&&&&&&&&\\[-10pt]
id     & $R_{\rm e}$ & $dR_{\rm e}$ & s. & $n$   & $ dn$    & s. & $\sigma_{\rm e}$ & +$d\sigma_{\rm e}$ & -$d\sigma_{\rm e}$ & s. & $\sigma_0$ & +$d\sigma_0$ & -$d\sigma_0$ & s. & $v_{\rm rot}$  & $+dv_{\rm rot}$ & $-dv_{\rm rot}$ & s. & $M_{\rm dyn}$  & $dM_{\rm dyn}$ & s. \\
       &  $[{\rm pc}]$       &  $[{\rm pc}]$          &        &       &         &          &  $[{\rm km/s}]$         &  $[{\rm km/s}]$         &   $[{\rm km/s}]$  &    &  $[{\rm km/s}]$         &  $[{\rm km/s}]$         &   $[{\rm km/s}]$     &       & $[{\rm km/s}]$       & $[{\rm km/s}]$   & $[{\rm km/s}]$  &          & $[\lg(M_{\odot})]$ & $[\lg(M_{\odot})]$ &        \\
\hline
$ 501$ & $  1879.9$ & $  188.0$ & $   1$ & $    2.8$ & $   0.4$ & $   1$ & $   75.7$ & $   3.8$ & $   3.8$ & $   1$ & $   76.7$ & $   3.8$ & $   3.8$ & $   1$ & $   58.3$ & $  -9.9$ & $   -9.9$ & $   1$ & $  9.75$ & $  0.09$ & $   1$\\
$ 502$ & $  2433.1$ & $  243.3$ & $   1$ & $    4.2$ & $   1.7$ & $   1$ & $   82.0$ & $   4.1$ & $   4.1$ & $   1$ & $   91.8$ & $   4.6$ & $   4.6$ & $   1$ & $   21.9$ & $  -9.9$ & $   -9.9$ & $   1$ & $  9.70$ & $  0.09$ & $   1$\\
$ 503$ & $  2043.4$ & $  204.3$ & $   1$ & $    8.4$ & $   2.0$ & $   1$ & $  128.5$ & $   6.4$ & $   6.4$ & $   1$ & $  132.4$ & $   6.6$ & $   6.6$ & $   1$ & $   36.2$ & $  -9.9$ & $   -9.9$ & $   1$ & $ 10.23$ & $  0.09$ & $   1$\\
$ 504$ & $  6949.9$ & $   -9.9$ & $   2$ & $    5.2$ & $   2.2$ & $   0$ & $  267.3$ & $  -9.9$ & $  -9.9$ & $ 101$ & $  290.4$ & $  -9.9$ & $  -9.9$ & $   2$ & $   -9.9$ & $  -9.9$ & $   -9.9$ & $   0$ & $ 11.68$ & $-99.99$ & $ 101$\\
$ 505$ & $  2393.8$ & $  239.4$ & $   1$ & $    4.4$ & $   0.4$ & $   1$ & $   62.2$ & $   3.1$ & $   3.1$ & $   1$ & $   56.2$ & $   2.8$ & $   2.8$ & $   1$ & $    8.5$ & $  -9.9$ & $   -9.9$ & $   1$ & $  9.65$ & $  0.09$ & $   1$\\
$ 506$ & $  1012.4$ & $  101.2$ & $   1$ & $    1.6$ & $   0.0$ & $   1$ & $  126.2$ & $   6.3$ & $   6.3$ & $   1$ & $  113.5$ & $   5.7$ & $   5.7$ & $   1$ & $   93.9$ & $  -9.9$ & $   -9.9$ & $   1$ & $ 10.19$ & $  0.09$ & $   1$\\
$ 507$ & $  1292.5$ & $  129.3$ & $   1$ & $    3.4$ & $   0.6$ & $   1$ & $   60.4$ & $   3.0$ & $   3.0$ & $   1$ & $   60.3$ & $   3.0$ & $   3.0$ & $   1$ & $   17.5$ & $  -9.9$ & $   -9.9$ & $   1$ & $  9.55$ & $  0.09$ & $   1$\\
$ 508$ & $  1362.4$ & $  136.2$ & $   1$ & $    0.8$ & $   0.1$ & $   1$ & $   66.4$ & $   3.3$ & $   3.3$ & $   1$ & $   63.4$ & $   3.2$ & $   3.2$ & $   1$ & $   22.5$ & $  -9.9$ & $   -9.9$ & $   1$ & $  9.54$ & $  0.09$ & $   1$\\
$ 509$ & $    96.3$ & $    6.1$ & $  10$ & $    0.6$ & $   0.2$ & $   0$ & $    6.7$ & $   1.4$ & $   1.4$ & $  12$ & $    6.7$ & $   1.4$ & $   1.4$ & $ 101$ & $    0.0$ & $  -9.9$ & $   -9.9$ & $   4$ & $  6.90$ & $  0.18$ & $ 101$\\
$ 510$ & $    91.5$ & $    9.7$ & $  10$ & $    0.5$ & $   0.2$ & $   0$ & $   92.0$ & $   5.0$ & $   5.0$ & $  12$ & $   92.0$ & $   5.0$ & $   5.0$ & $ 101$ & $   55.0$ & $  -9.9$ & $   -9.9$ & $   4$ & $  9.15$ & $  0.07$ & $ 101$\\
&&&&&&&&&&&&&&&&&&&&&&& \\[-10pt]
\hline
\end{tabular}
\end{table}
\end{landscape}

\section[Stellar populations]{Properties of stellar populations}
\label{sec:properties}

\subsection{General procedures}
\label{sec:properties1}

In Tab.~(\ref{tab52}), we list some properties of the stellar populations of the galaxies, namely total masses of the stellar populations, and their SSP-equivalent ages and metallicities.

\subsubsection{Metallicities}
\label{sec:metallicities}

If no individual estimate for metallicity of a galaxy is available in the literature, it is estimated here based on the SSP-models from \citet{Bruzual2003}. This estimate of the metallicities of the galaxies is done in three steps.

The first step is to find interpolation formulae that quantify how metallicity is connected to colour according to the SSP-models by \citet{Bruzual2003}. This is done by fitting functions of the form 
\begin{equation}
\rm{[Z/H]}_{\rm SSP}=a \exp[-(X-Y)]+b\, (X-Y)+c
\label{eq:MetalFit}
\end{equation}
to data from SSP-models by \citet{Bruzual2003}, where $\rm{[Z/H]}_{\rm SSP}$ is the logarithm of the metallicity in Solar units and $(X-Y)$ is the colour of the SSP. The best fitting parameters $a$, $b$ and $c$ for different ages and colours are listed in Tab.~(\ref{tab:MetalFit}). The fits to the SSP-models are shown in Fig.~(\ref{fig:MetalFit}). If no individual estimate for the SSP-equivalent age of a galaxy is available, a value from Tab.~(\ref{tab:means}) is assigned to it based on its $V$-band luminosity.

The second step is to estimate  $\rm{[Z/H]}_{\rm SSP}$ of the galaxy with Eq.~(\ref{eq:MetalFit}).

The third step is to compare the estimates for ${\rm [Z/H]}_{\rm SSP}$ with the previously published values for all galaxies where both an estimate with Eq.~(\ref{eq:MetalFit}) is possible and a previously published values is available. This is done by evaluating the equation
\begin{equation}
\Delta {\rm [Z/H]}^i={\rm [Z/H]}_{\rm pub}^{i}-{\rm [Z/H]}_{\rm SSP}^{i},
\label{eq:gaugeMetal}
\end{equation}
where ${\rm [Z/H]}_{\rm pub}^{i}$ is the previously published value for the [Z/H] of the $i$th galaxy, and ${\rm [Z/H]}_{\rm SSP}^{i}$ is the [Z/H] of the same galaxy calculated with Eq.~(\ref{eq:MetalFit}). The median value of the results from Eq.~(\ref{eq:gaugeMetal}), $\Delta {\rm [Z/H]}$, is then added to the estimates ${\rm [Z/H]}_{\rm SSP}^{i}$, in order to gauge them to the previously published values. When $(g-r)$ is available in our catalogue, $(g-i)$ is available from the same source, while the same is usually not true for $(B-V)$ and $(V-I)$. We therefore use the the mean of the estimates based on $(g-r)$ and $(g-i)$ for the final estimates of $[Z/H]$ from colours in the SDSS-system, while we use either $(B-V)$ or $(V-I)$ for estimates based on colours in the standard system.

The gauging is illustrated in Fig.~(\ref{fig:MetalscatterBV}) for estimates of [Z/H] from $(B-V)$, where we find $\Delta {\rm [Z/H]}=0.24$. If the values for ${\rm [Z/H]}_{\rm SSP}^{i}$  in eq.~(\ref{eq:gaugeMetal}) are calculated as a mean of the estimates based on $(g-r)$ and $(g-i)$ instead of a single estimate based on $(B-V)$, we find $\Delta {\rm [Z/H]}=0.04$. Due to the smaller value for $\Delta {\rm [Z/H]}$, we prefer estimates of [Z/H] based on $(g-r)$ and $(g-i)$ over those from $(B-V)$. We abstain from using estimates for the metallicity based on $(V-I)$, since for those estimates $\Delta {\rm [Z/H]}$ is very large.

There are several possible reasons why this third step is necessary. The first reason is that SSP-models are used here for simplicity, while the actual stellar populations of the galaxies are often much more complex than a SSP. The second reason is that the SSP-equivalent ages and metallicities from the literature were often derived from SSP-models other than the ones by \citet{Bruzual2003}, which we use as a reference. This may lead to systematic errors. We nevertheless prefer using the SSP-models by \citet{Bruzual2003} in all cases, since they do not only cover magnitudes in the standard system, but also magnitudes in the AB-magnitude system, which is very similar to the SDSS-magnitude system. (In fact, the SDSS-magnitude system was designed to reproduce the AB-magnitude system). This is a desirable feature of the SSP-models by \citet{Bruzual2003}, since the catalogue presented here includes a vast sample of magnitudes from the SDSS, which are by nature internally consistent. Moreover, there is a large number of galaxies in our catalog, for which magnitudes are available only from the SDSS. The third reason is that the best estimates for the ages of the galaxies are mapped to a rather coarse grid with only six different options for the best estimate for the age of the galaxy.

The uncertainty of the metallicities calculated by the method described above is estimated as the standard deviation from the relation
\begin{equation}
{\rm [Z/H]}_{\rm SSP}+\Delta {\rm [Z/H]}={\rm [Z/H]}_{\rm pub},
\label{eq:scatterMetal}
\end{equation}
where $\Delta {\rm [Z/H]}$ is the median of the values calculated with Eq.~(\ref{eq:gaugeMetal}).

\begin{figure}
\centering
\includegraphics[scale=0.78]{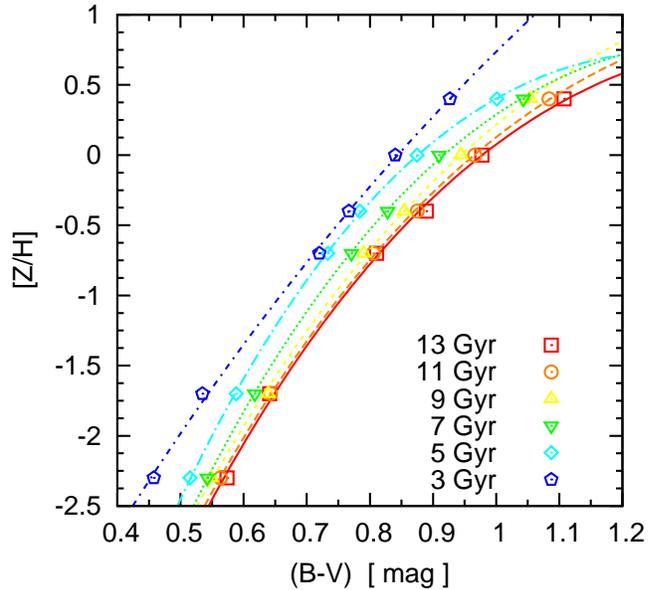}
\caption{\label{fig:MetalFit} The interpolations for the metallicities as functions of $(B-V)$-colours for different ages. The curves show the best fit of Eq.~(\ref{eq:MetalFit}) to the data from a SSP-model from \citet{Bruzual2003} for a 13~Gyr old, a 11~Gyr old, a 9~Gyr old, a 7~Gyr old, a 5~Gyr old, and a 3~Gyr old SSP from top to bottom. The symbols with the same colour as the interpolation function show the data to which the according curve was fitted. Figures as this one, but for other colours, show similar characteristics.}
\end{figure}

\begin{figure*}
\centering
\includegraphics[scale=0.78]{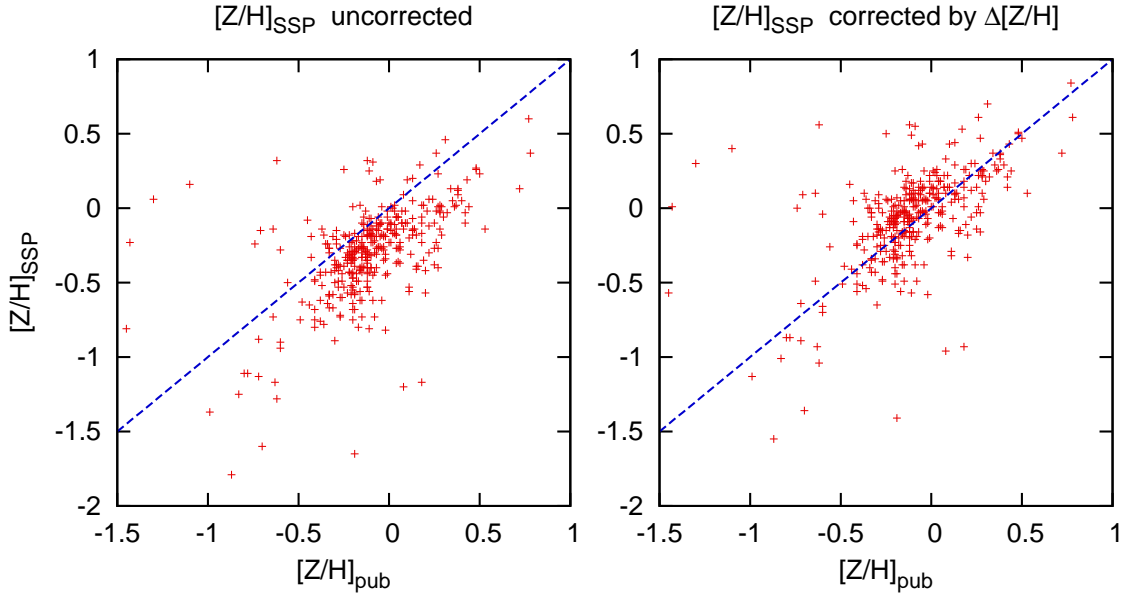}
\caption{\label{fig:MetalscatterBV} The gauging of $[Z/H]$ in order to obtain the best possible agreement between the values for $[Z/H]$ in the literature on the x-axis and the values for [Z/H] calculated from SSP-models on the y-axis. In the left panel, the estimates based on the SSP-models have been calculated with Eq.~\ref{eq:MetalFit} using $(B-V)$. In the right panel, the estimates from the SSP-models have additionally been gauged to the observed values using Eqs.~(\ref{eq:gaugeMetal}) and~(\ref{eq:scatterMetal}). If $[Z/H]$ is calculated from $(g-r)$ and $(g-i)$ instead of $(B-V)$, the correction shown in this figure for $(B-V)$ is much smaller, and negligible over the scatter of the data. For this reason, values for $[Z/H]$ calculated from $(g-r)$ and $(g-i)$ are preferred over those calculated from $(B-V)$, if both values for $[Z/H]$ are available.}
\end{figure*}

\subsubsection{Masses of the stellar populations}
\label{sec:stellarmass}

For most extragalactic objects, the mass of their stellar populations ($M_{\rm s}$) cannot be estimated directly from star counts. Quite often, the $M_{\rm s}$ of such galaxies is estimated by simply assuming a fixed mass-to-light ratio ($M_{\rm s}/L$-ratio) for all galaxies, and multiplying it by their respective luminosities. For our catalog, we attempt to make more precise estimates for $M_{\rm s}$ by using age and colour as additional parameters besides luminosity. The different combinations of age and colours considered here allow the $V$-band $M_{\rm s}/L$-ratio to assume values between $0.6 \, {\rm M_{\odot}}/{\rm L_{\odot}}$ and $6 \, {\rm M_{\odot}}/{\rm L_{\odot}}$.

In practice, the $M_{\rm s}/L_V$-ratio of a given galaxy is calculated from interpolation formulae of the form
\begin{equation}
 \frac{M}{L_{V}}=a \, \arctan [ b \, (X-Y) + c ] + d,
\label{eq:MLFit}
\end{equation}
which are fitted to data from the SSP-models by \citet{Bruzual2003}. In this equation, $(X-Y)$ is the colour of a galaxy with apparent magnitudes $m_X$ and $m_Y$. The best-fitting parameters $a$, $b$, $c$ and $d$ in Eq.~(\ref{eq:MLFit}) for different colours and ages are shown in Tab.~(\ref{tab:MLFit}). The data from the SSP-models and the fits to them are shown in Fig.~(\ref{fig:MLfit}).

For many galaxies, no individual estimates for their age is given, and for some neither an estimate for their age nor for their colour is available. We nevertheless provide rough estimates for their $M_{\rm s}$ by setting the unknown ages and colours to mean values that are calculated from individual ages and colours of galaxies with similar $V$-band luminosities. These mean values, and the number of galaxies from which they are estimated, are summarised in Tab.~(\ref{tab:means}).

Gauging our estimates for $M_{\rm s}$ to independent estimates for $M_{\rm s}$ as done for metallicities (Section~\ref{sec:metallicities}) is not possible, since there are only very few direct estimates of $M_{\rm s}$ of galaxies in the literature. Moreover, this independent data for $M_{\rm s}$ is only available for very specific galaxies, namely dwarf galaxies in the Local Group, which are all at the faint end of the luminosity function for ETGs.

We prefer estimates for $M_{\rm s}$ based on $(g-r)$ and $(g-i)$ over those based only on $(B-V)$, and abstain from using $(V-I)$ for estimates of $M_{\rm s}$. This is motivated on the findings on estimates of the metallicities of the galaxies, where we used the same SSP-models and a similar method (Section~\ref{sec:metallicities}).

\begin{figure}
\centering
\includegraphics[scale=0.78]{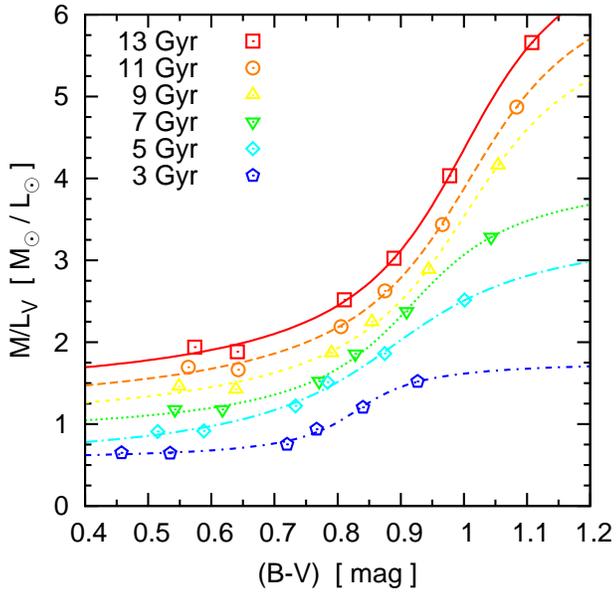}
\caption{\label{fig:MLfit} The interpolations for the $M_{\rm s}/L_V$-ratios as functions of $(B-V)$-colours for different ages. The curves show the best fit of Eq.~(\ref{eq:MLFit}) to the data from a SSP-model from \citet{Bruzual2003} for a 13~Gyr old, a 11~Gyr old, a 9~Gyr old, a 7~Gyr old, a 5~Gyr old, and a 3~Gyr old SSP from top to bottom. The symbols with the same colour as the interpolation function show the data to which the according curve was fitted. Figures as this one, but for other colours, show similar characteristics.}
\end{figure}

\begin{landscape}
\begin{table}
\caption{\label{tab:MetalFit} The best-fitting parameters in Eq.~(\ref{eq:MetalFit}) for different colours in the standard system and different ages. Column~1 lists the age of the SSP to which Eq.~(\ref{eq:MetalFit}) is fitted. Columns~2,~3 and~4, respectively, show the best-fitting parameters $a$, $b$ and $c$ if $(U-B)$ is calculated from the metallicity an the age of the galaxy. This is repeated for $(B-V)$ in columns~5,~6 and~7 and for $(V-I)$ in columns~8,~9 and~10. The parameters found for a 13~Gyr old SSP are used for galaxies with ages $t>12 \ {\rm Gyr}$, the ones found for a 11~Gyr old SSP are used for galaxies with $10 \ {\rm Gyr} < t \le 12 \ {\rm Gyr}$, the ones found for a 9~Gyr old SSP are used for galaxies with $8 \ {\rm Gyr} < t \le 10 \ {\rm Gyr}$, the ones found for a 7~Gyr old SSP are used for galaxies with $6 \ {\rm Gyr} < t \le 8 \ {\rm Gyr}$, the ones found for a 5~Gyr old SSP are used for galaxies with $4 \ {\rm Gyr} < t \le 6 \ {\rm Gyr}$, and the ones found for a 3~Gyr old SSP are used for galaxies with $t \le 4 \ {\rm Gyr}$. It seems unnecessary to consider SSP-models for even younger populations, since there are hardly any galaxies which have stellar populations with SSP-equivalent ages less than 2~Gyr.}
\centering
\vspace{2mm}
\begin{tabular}{lrrrrrrrrrrrr}
\hline
&&&&&&&&&&&&\\[-10pt]
age       & $a_{(B-V)}$ & $b_{(B-V)}$ & $c_{(B-V)}$  & $a_{(V-I)}$  &  $b_{(V-I)}$  & $c_{(V-I)}$  & $a_{(g-r)}$  &  $b_{(g-r)}$  & $c_{(g-r)}$  & $a_{(g-i)}$ & $b_{(g-i)}$ & $c_{(g-i)}$ \\
\hline
13 Gyr  & $-23.5832$  &  $-5.3339$  &  $14.0585$   & $-36.9288$ &   $-7.3174$  &  $19.7142$ & $-27.2662$ &    $-7.9619$ &  $18.4040$ & $-17.6094$ & $-2.7110$ &   $8.3964$ \\
11 Gyr  & $-21.7337$  &  $-4.4787$  &  $12.6014$   & $-37.6461$ &   $-7.5559$  &  $20.2622$ & $-25.9793$ &    $-7.3407$ &  $17.3708$ & $-17.4673$ & $-2.6669$ &   $8.3451$ \\
9  Gyr   & $-20.1594$  &  $-3.7445$  &  $11.3780$   & $-41.9067$ &   $-8.9716$  &  $23.3369$ & $-25.1418$ &    $-6.9482$ &  $16.7607$ & $-18.2492$ & $-2.9573$ &   $9.0140$ \\
7  Gyr   & $-26.5008$  &  $-6.6921$  &  $16.7299$   & $-50.2784$ & $-11.8803$  &  $29.4590$ & $-31.1215$ &  $-10.0752$ &  $22.0629$ & $-21.9729$ & $-4.3914$ & $11.9625$ \\
5  Gyr   & $-28.8649$  &  $-8.0716$  &  $19.0924$   & $-58.3345$ & $-15.2585$  &  $36.1169$ & $-32.5321$ &  $-11.2268$ &  $23.7681$ & $-22.4748$ & $-4.9813$ & $12.9636$ \\
3  Gyr   & $  -9.0532$  &  $ 1.1180$   &  $22.4484$   & $-38.2278$ &   $-8.8094$ &  $22.4484$ &   $-9.7218$ &     $0.5517$ &    $4.3199$ & $-10.5927$ & $-0.6135$ &   $4.3504$ \\
 
&&&&&&&&&&&& \\[-10pt]
\hline
\end{tabular}
\end{table}

\begin{table}
\caption{\label{tab:MLFit} The best-fitting parameters in Eq.~(\ref{eq:MLFit}) for different colours and different ages. Column~1 lists the age of the SSP to which Eq.~(\ref{eq:MLFit}) is fitted. Columns~2,~3,~4 and~5, respectively, show the best-fitting parameters $a$, $b$, $c$ and $d$ if $M_{\rm s}/L_V$ is calculated from $(B-V)$ and the age of the galaxy. This is repeated for $(V-I)$ in columns~6,~7~8 and~9, for $(g-r)$ in columns~10,~11~12 and~13 and for $(g-i)$ in columns~14,~15~16 and~17. The parameters found for a 13~Gyr old SSP are used for galaxies with ages $t>12 \ {\rm Gyr}$, the ones found for a 11~Gyr old SSP are used for galaxies with $10 \ {\rm Gyr} < t \le 12 \ {\rm Gyr}$, the ones found for a 9~Gyr old SSP are used for galaxies with $8 \ {\rm Gyr} < t \le 10 \ {\rm Gyr}$, the ones found for a 7~Gyr old SSP are used for galaxies with $6 \ {\rm Gyr} < t \le 8 \ {\rm Gyr}$, the ones found for a 5~Gyr old SSP are used for galaxies with $4 \ {\rm Gyr} < t \le 6 \ {\rm Gyr}$, and the ones found for a 3~Gyr old SSP are used for galaxies with $t \le 4 \ {\rm Gyr}$. It seems unnecessary to consider SSP-models for even younger populations, since there are hardly any galaxies which have stellar populations with SSP-equivalent ages less than 2~Gyr.}
\centering
\vspace{2mm}
\begin{tabular}{lrrrrrrrrrrrrrrrr}
\hline
&&&&&&&&&&&&&&&& \\[-10pt]
age    & $a_{(B-V)}$ &  $b_{(B-V)}$   & $c_{(B-V)}$ & $d_{(B-V)}$ &  $a_{(V-I)}$   & $b_{(V-I)}$ & $c_{(V-I)}$ &  $d_{(V-I)}$   & $a_{(g-r)}$  & $b_{(g-r)}$ &  $c_{(g-r)}$   & $d_{(g-r)}$  & $a_{(g-i)}$  & $b_{(g-i)}$ &  $c_{(g-i)}$   & $d_{(g-i)}$ \\
\hline
13 Gyr     &  $1.9779$ & $7.1943$  & $-7.1914$ & $4.3499$ & $1.5027$ & $11.4859$ & $-14.0347$ & $3.9112$ & $1.8220$ & $7.9883$ & $-6.8406$ & $ 4.1377$ & $1.6325$ & $6.3431$ & $-7.8626$ & $3.9850$ \\
11 Gyr     &  $1.8771$ & $6.8388$  & $-6.8937$ & $3.9783$ & $1.3088$ & $10.9951$ & $-13.3228$ & $3.3926$ & $1.6342$ & $7.7330$ & $-6.6032$ & $ 3.6605$ & $1.4385$ & $6.0948$ & $-7.4920$ & $3.4812$ \\
9  Gyr      &  $1.7431$ & $6.8583$  & $-6.8881$ & $3.5863$ & $1.1431$ & $10.5101$ & $-12.5369$ & $2.9105$ & $1.4285$ & $7.8588$ & $-6.5935$ & $ 3.1746$ & $1.2596$ & $5.9876$ & $-7.2125$ & $3.0023$ \\
7  Gyr      &  $1.0694$ & $7.6960$  & $-7.0625$ & $2.4626$ & $0.8449$ & $11.7422$ & $-13.4595$ & $2.2627$ & $0.9590$ & $9.3954$ & $-7.2787$ & $ 2.3436$ & $0.8942$ & $7.0160$ & $7.8888$ & $2.2909$ \\
5  Gyr      &  $0.9661$ & $5.6806$  & $-5.0505$ & $1.9641$ & $0.8088$ & $6.1142$   & $-6.6885$   & $1.7338$ & $0.8227$ & $7.1420$ & $-5.2455$ & $ 1.8194$ & $0.8073$ & $4.5046$ & $4.7658$ & $1.7730$ \\
3  Gyr      &  $0.3948$ & $12.6401$ & $-10.5005$ & $1.1689$ & $0.4355$ & $8.8163$ & $-9.6147$  & $1.1762$ & $0.3872$ & $13.8092$ & $-9.7617$ & $1.1601$ & $0.4031$ & $7.6375$ & $7.8725$ & $1.1630$ \\
 
&&&&&&&&&&&&&&&& \\[-10pt]
\hline
\end{tabular}
\end{table}
\end{landscape}

\begin{table*}
\caption{\label{tab:means} Mean values for ages and colours of galaxies in different luminosity intervals. Column~1 lists the luminosity interval that is considered. Column~2 shows the mean SSP-equivalent ages of galaxies in the luminosity intervals specified in column~1. Column~3 lists the number of galaxies that were used for calculating the mean ages. Column~4 shows the mean $(B-V)$ colours for the luminosity intervals listed in column~1. Column~5 lists the number of galaxies that were used for calculating the mean $(B-V)$. Column~6 shows the mean $(V-I)$ colours for the luminosity intervals listed in column~1. Column~7 lists the number of galaxies that were used for calculating the mean $(V-I)$. Column~8 shows the mean $(g-r)$ colours for the luminosity intervals listed in column~1. Column~9 lists the number of galaxies that were used for calculating the mean $(g-r)$. Column~10 shows the mean $(g-i)$ colours for the luminosity intervals listed in column~1. Column~11 lists the number of galaxies that were used for calculating the mean $(g-i)$.}
\centering
\vspace{2mm}
\begin{tabular}{lrrrrrrrrrr}
\hline
&&&&&&&&&&\\[-10pt]
$L_V$                      & $t$           & $N_t$ & $(B-V)$ & $N_{\rm (B-V)}$ & $(V-I)$ & $N_{\rm (V-I)}$ & $(g-r)$ & $N_{\rm (g-r)}$ & $(g-i)$ & $N_{\rm (g-i)}$   \\
$\log_{10} \left(\frac{L_V}{{\rm L}_{\odot}}\right)$ & Gyr & & $[{\rm mag}]$ &       & $[{\rm mag}]$     &                          & $[{\rm mag}]$     &                          & $[{\rm mag}]$     &                           \\
\hline
$L_V>10.7           $ & $10.69 $  & $36$   & $0.92$   & $60$                  & $1.10$ &     $7$               & $0.83$ &   $35$               & $1.22$ &   $35$                \\
$9.7<L_V \le 10.7$ & $  9.34 $  & $181$ & $0.86$   & $271$                & $1.08$ &   $69$               & $0.78$ & $212$               & $1.07$ & $211$                \\
$8.0<L_V \le   9.7$ & $  7.56 $  & $59$   & $0.82$   & $127$                & $0.91$ & $146$               & $0.68$ & $120$               & $1.00$ & $120$                \\
$L_V \le 8.0         $ & $  8.34 $  & $29$   & $0.70$   & $67$                  & $0.88$ & $392$               & $0.53$ &   $89$               & $0.75$ &   $89$                \\

&&&&&&&&&&\\[-10pt]
\hline
\end{tabular}
\end{table*}

\subsection{Detailed description of the data}
\label{sec:properties2}

Below, we explain the entries in Tab.~(\ref{tab52}) column by column.

{\bf Column~1} lists the running number of the galaxy.

{\bf Column~2} shows $t$, the SSP-equivalent age of the stellar populations of the galaxies. Note that the real stellar populations of ETGs can deviate quite substantially from being a SSP, especially at low luminosities. The SSP-equivalent age of a galaxy is, however, a convenient way to characterise the typical ages of the stars in a galaxy with a single number. The entry in column~2 is set to $-9.9$ if $t$ is unknown.

{\bf Column~3} lists the positive uncertainties of $t$. The entry in that column is set to $-9.9$ if this uncertainty is unknown.

{\bf Column~4} lists the absolute values of the negative uncertainties of $t$, since the uncertainties of $t$ are asymmetric for some galaxies. The entry in that column is set to $-9.9$, if this uncertainty is unknown.

{\bf Column~5}  lists the sources for the estimates of $t$ and their uncertainties. The number in this column is set to `0' if no data on $t$ is available. The number is set to `1' if the galaxy is included in the ATLAS$^{\rm 3D}$ survey, and that the quoted value for $t$ is taken from \citet{McDermid2015}. The number is set to `2' if $t$ is taken from \citet{Thomas2005}, to `3' if the galaxy is part of the Coma Cluster and its $t$ is taken from \citet{Matkovic2009}, to `4' if the galaxy is part of the Coma Cluster and its $t$ is taken from \citet{SanchezBlazquez2006}, to `5' if the galaxy is part of the cluster Abell~496 and its $t$ is taken from \citet{Chilingarian2008}, and to `6' if the estimate for $t$ is based on data on Local-Group ETGs from \citet{Weisz2014}. As a proxy for the $t$ of the ETGs from \citet{Weisz2014}, we list the times at which 70 percent of their stellar populations have formed according to them. The number is set to `7' if the galaxy is part of the Virgo Cluster and its $t$ is taken from \citet{Toloba2014}, and to `8' if the galaxy is part of the Virgo Cluster and its $t$ is taken from \citet{Guerou2015}.

{\bf Column~6} shows [Z/H], which is the logarithm to the base of 10 of the SSP-equivalent metal abundance of the galaxies in Solar units. As with the SSP-equivalent age $t$, also the SSP-equivalent metallicity is a number that captures the general properties of the actual stellar population of a galaxy in a single number, while the actual stellar populations can be quite complex.

The entry in column~6 is set to $99.99$ if [Z/H] is unknown.

{\bf Column~7} lists the positive uncertainties of [Z/H]. The entry in that column is set to $99.99$ if this uncertainty is unknown.

{\bf Column~8} lists the absolute values of the negative uncertainties of [Z/H], since the uncertainties of [Z/H] are asymmetric for some galaxies. The entry in that column is set to $99.99$, if this uncertainty is unknown.

{\bf Column~9} lists the sources for the estimates of [Z/H] and their uncertainties. The number in this column is set to `0' if no data on [Z/H] is available. The number is set to `1' if the galaxy is included in the ATLAS$^{\rm 3D}$ survey, and the quoted value for [Z/H] is taken from \citet{McDermid2015}, to `2' if [Z/H] is taken from \citet{Thomas2005}, to `3' if the galaxy is part of the Coma Cluster and its [Z/H] is taken from \citet{Matkovic2009}, to `4' if the galaxy is part of the Coma Cluster and its [Z/H] is taken from \citet{SanchezBlazquez2006}, to `5' if the galaxy is part of the cluster Abell~496 and its [Z/H] is taken from \citet{Chilingarian2008}, and to `6' if the galaxy is part of the Local Group and its [Z/H] is based on the values from \citet{McConnachie2012}. We note that \citet{McConnachie2012} list [Fe/H] instead of [Z/H], but we leave the values taken from \citet{McConnachie2012} for the metallicity unchanged; i.e. we implicitly assume Solar abundance ratios in the galaxies in the Local Group. The number is set to `7' if the galaxy is part of the Virgo Cluster and its [Z/H] is taken from \citet{Toloba2014}, and to `8' if the galaxy is part of the Virgo Cluster and its [Z/H] is taken from \citet{Guerou2015}. The number is set to `1011' if $(g-r)$, $(g-i)$ and an individual estimate for $t$ of the galaxy are available. Its [Z/H] is then estimated on this basis, using Eq.~(\ref{eq:MetalFit}) with the appropriate entries in Tab.~(\ref{tab:MetalFit}). The number is set to `1010' if $(g-r)$, $(g-i)$ and $L_V$ of the galaxy are available, but $t$ is unknown for it. In this case, its $t$ is assumed to be equal to the mean $t$ of galaxies with similar $L_V$, as listed in Tab.~(\ref{tab:means}). Its [Z/H] is then estimated using Eq.~(\ref{eq:MetalFit}) with the appropriate entries in Tab.~(\ref{tab:MetalFit}). Whenever $[Z/H]$ is calculated from $(g-r)$ and $(g-i)$, the quoted value is the mean of the estimates from these two colours. The number is set to `1021' if $(B-V)$, $L_V$ and an individual estimate for $t$ of the galaxy are available, but $(g-r)$ and $(g-i)$ are unknown. Its [Z/H] is then estimated on this basis, using Eq.~(\ref{eq:MetalFit}) with the appropriate entries in Tab.~(\ref{tab:MetalFit}). The number is set to `1020' if $(B-V)$ and $L_V$ of the galaxy are available, but its $(g-r)$, $(g-i)$ and $t$ are unknown. In this case, its $t$ is assumed to be equal to the mean $t$ of galaxies with similar $L_V$, as listed in Tab.~(\ref{tab:means}). Its [Z/H] is then estimated on this basis, using Eq.~(\ref{eq:MetalFit}) with the appropriate entries in Tab.~(\ref{tab:MetalFit}). Fig.~(\ref{fig:Zpublished}) shows a comparison between values for [Z/H] taken from the literature and the estimates for [Z/H] done for this catalog, if no published value was found. It is apparent from Fig.~(\ref{fig:Zpublished}) that for $L_V<10^9 \, {\rm M}_{\odot}$, the scatter of the values for [Z/H] calculated for this catalogue using Eq.~(\ref{eq:MetalFit}) is much larger than the scatter of the values obtained from the literature. However, the estimates for the metallicities calculated here still tend to gather at the same regions in the luminosity-metallicity plane as the metallicity estimates from the literature. Nevertheless, the scatter implies that the values obtained for [Z/H] using Eq.~(\ref{eq:MetalFit}) are probably not very reliable for ETGs with $L_V<10^9 \, {\rm M}_{\odot}$. It is therefore probably the best to exclude such galaxies in applications, where accurate estimates for their metallicities are essential.

\begin{figure*}
\centering
\includegraphics[scale=0.78]{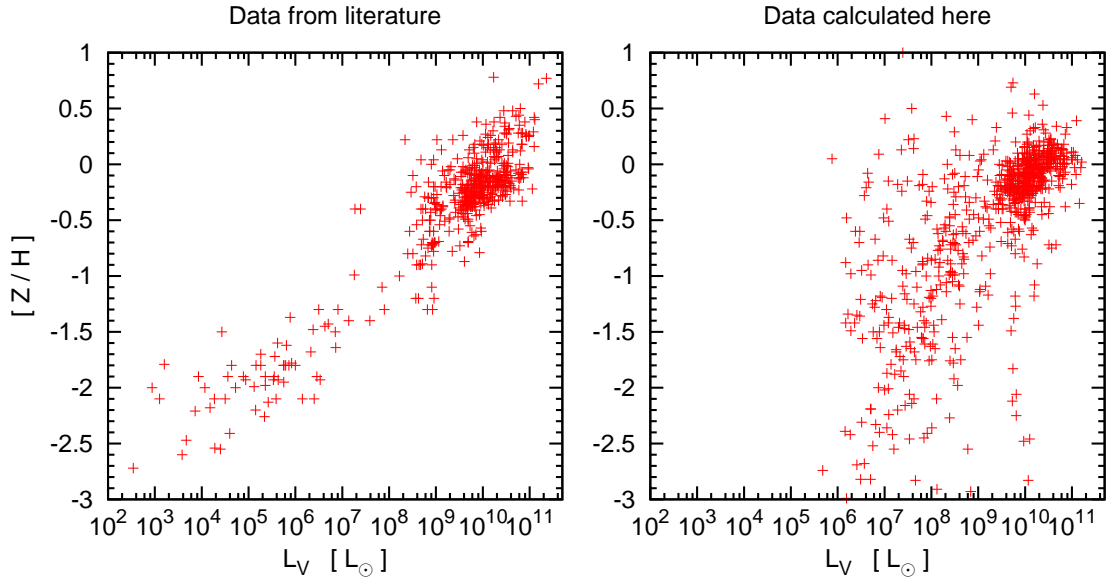}
\caption{\label{fig:Zpublished} Comparison between the values for [Z/H] taken from the literature and values for [Z/H] calculated for this catalogue using Eq.~(\ref{eq:MetalFit}), if no estimate from the literature was available. The left panel shows the data from the literature and the right panel shows the data calculated using Eq.~(\ref{eq:MetalFit}).}
\end{figure*}

{\bf Column~10} lists $M_{\rm s}$, the masses of the stellar population of the galaxies. The numbers actually shown in column~10 are the logarithms to the base of 10 of $M_{\rm s}$ in Solar units. If $M_{\rm s}$ cannot be estimated, the number in this column is set to $-99.99$. Fig.~(\ref{fig:L-MLs}) shows the $M_{\rm s}/L_V$-ratio of the ETGs in this catalogue over their $L_V$. The concentration of the data at a $M_{\rm s}/L_V$-ratio of about 1.5 ${\rm M_{\odot}}/{\rm L_{\odot}}$ illustrates that the colours of low-luminosity galaxies are often unknown, so that a typical average colour of known galaxies in the same luminosity range was assumed. High $M_{\rm s}/L_V$-ratios indicate that the galaxies have (measured) colours which imply that their stellar populations are old, metal-rich or both.

{\bf Column~11} shows the logarithmic uncertainty to the values of $M_{\rm s}$. The value in this column is set to $-99.99$, if the uncertainty of $M_{\rm s}$ could not be estimated.

{\bf Column~12} lists the sources for the estimates of $M_{\rm s}$ and their uncertainties. The number in this column is set to `0' if no data on $M_{\rm s}$ is available. The number is set to `1' if the galaxy is a member of the Local Group and its estimate for $M_{\rm s}$ is taken from table~1 in \citet{Misgeld2011}. The number is set to `1011' if $(g-r)$, $(g-i)$, $L_V$ and an individual estimate for $t$ are available for the galaxy. $M_{\rm s}$ is then estimated for it on this basis, using Eq.~(\ref{eq:MLFit}) with the appropriate entries in Tab.~(\ref{tab:MLFit}). A `1010' indicates that $(g-r)$, $(g-i)$ and $L_V$ of the galaxy are available, but $t$ is unknown for it. In this case, $t$ is assumed to be equal to the mean age of galaxies with similar $L_V$, as listed in Tab.~(\ref{tab:means}). $M_{\rm s}$ is then estimated for it on this basis, using Eq.~(\ref{eq:MLFit}) with the appropriate entries in Tab.~(\ref{tab:MLFit}). The number is set to `1021' if $(B-V)$, $L_V$ and an individual estimate for $t$ are available for the galaxy, but its $(g-r)$ and $(g-i)$ are unknown. Its $M_{\rm s}$ is then estimated on this basis, using Eq.~(\ref{eq:MLFit}) with the appropriate entries in Tab.~(\ref{tab:MLFit}). The number is set to `1020' if $(B-V)$ and $L_V$ of the galaxy are available, but its $(g-r)$, $(g-i)$ and $t$ are unknown. In this case, its age is assumed to be equal to the mean age of galaxies with similar $L_V$, as listed in Tab.~(\ref{tab:means}). Its $M_{\rm s}$ is then estimated on this basis, using Eq.~(\ref{eq:MLFit}) with the appropriate entries in Tab.~(\ref{tab:MLFit}). The number is set to `1000' if only $L_V$ of the galaxy is available, while its $(g-r)$, $(g-i)$, $(B-V)$ and $t$ are unknown. In this case, its $(g-r)$, $(g-i)$ and age is assumed to be equal to the mean $(g-r)$, $(g-i)$ and $t$ of galaxies with similar $L_V$, as listed in Tab.~(\ref{tab:means}). Its $M_{\rm s}$ is then estimated on this basis, using Eq.~(\ref{eq:MLFit}) with the appropriate entries in Tab.~(\ref{tab:MLFit}). Whenever $[Z/H]$ is calculated from $(g-r)$ and $(g-i)$, the quoted value is the mean of the estimates from these two colours.

\begin{figure}
\centering
\includegraphics[scale=0.78]{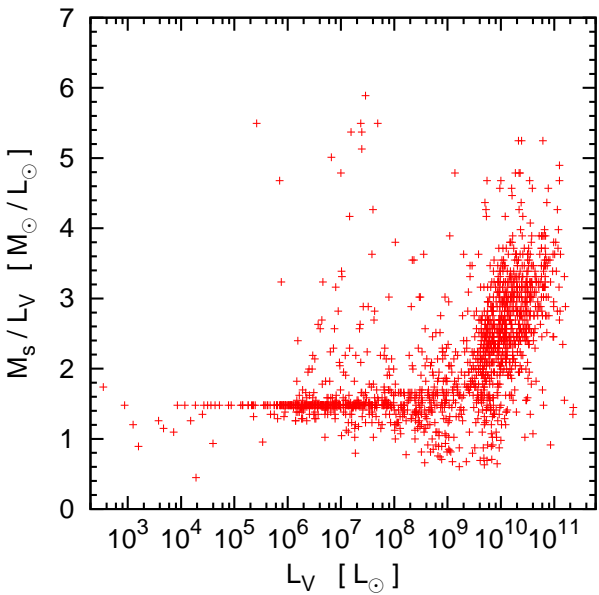}
\caption{\label{fig:L-MLs} The ratio between the mass of the stellar population and luminosity, $M_{\rm s}/L_V$, over $L_V$ for the ETGs in this catalogue.}
\end{figure}

\begin{table*}
\caption{\label{tab52} Stellar populations of the galaxies in this catalogue. A portion of the Table is shown here for guidance regarding its contents and form. A detailed description of the contents of this table is given in Section~(\ref{sec:properties2}). The table in its entirety can be downloaded at http://www.astro-udec.cl/mf/catalogue/}
\centering
\vspace{2mm}
\begin{tabular}{rrrrrrrrrrrr}
\hline
&&&&&&&&&&&\\[-10pt]
id   & $t$ & $+dt$ & $-dt$ & s. & [Z/H] & $+d$[Z/H] &$-d$[Z/H] & s. & $M_{\rm s}$ & $dM_{\rm s}$ & s.\\
     & $[{\rm Gyr}]$ & $[{\rm Gyr}]$    & $[{\rm Gyr}]$   &             &          &                  &                &              & $\lg(M_{\odot})$       & $\lg(M_{\odot})$         &         \\
\hline
$ 501$ & $  3.5$ & $  0.6$ & $  0.6$ & $   1$ & $ -0.38$ & $  0.06$ & $  0.06$ & $   1$ & $  9.81$ & $-99.99$ & $1011$\\
$ 502$ & $  1.0$ & $  0.1$ & $  0.1$ & $   1$ & $  0.18$ & $  0.06$ & $  0.06$ & $   1$ & $  9.68$ & $-99.99$ & $1011$\\
$ 503$ & $  9.7$ & $  1.6$ & $  1.6$ & $   1$ & $ -0.36$ & $  0.05$ & $  0.05$ & $   1$ & $ 10.09$ & $-99.99$ & $1011$\\
$ 504$ & $ -9.9$ & $ -9.9$ & $ -9.9$ & $   0$ & $  0.06$ & $ 99.99$ & $ 99.99$ & $1020$ & $ 11.46$ & $-99.99$ & $1020$\\
$ 505$ & $  2.5$ & $  0.3$ & $  0.3$ & $   1$ & $ -0.78$ & $  0.06$ & $  0.06$ & $   1$ & $  9.18$ & $-99.99$ & $1011$\\
$ 506$ & $  9.3$ & $  1.7$ & $  1.7$ & $   1$ & $ -0.36$ & $  0.05$ & $  0.05$ & $   1$ & $ 10.18$ & $-99.99$ & $1011$\\
$ 507$ & $  3.5$ & $  0.4$ & $  0.4$ & $   1$ & $ -0.87$ & $  0.08$ & $  0.08$ & $   1$ & $  9.45$ & $-99.99$ & $1011$\\
$ 508$ & $  2.3$ & $  0.3$ & $  0.3$ & $   1$ & $ -0.36$ & $  0.05$ & $  0.05$ & $   1$ & $  9.62$ & $-99.99$ & $1011$\\
$ 509$ & $ -9.9$ & $ -9.9$ & $ -9.9$ & $   0$ & $ -2.47$ & $  0.06$ & $  0.06$ & $   6$ & $  3.73$ & $  0.04$ & $   1$\\
$ 510$ & $  6.2$ & $ -9.9$ & $ -9.9$ & $   6$ & $ -0.25$ & $ 99.99$ & $ 99.99$ & $   6$ & $  8.95$ & $-99.99$ & $1021$\\
&&&&&&&&&&& \\[-10pt]
\hline
\end{tabular}
\end{table*}

\section{Conclusion}
\label{sec:conclusion}

We have collected a catalogue of 1715 early-type galaxies from the literature, spanning the luminosity range from faint dwarf spheroidal galaxies to giant elliptical galaxies. The catalogue is presented in ten tables, which are dedicated to the names of the galaxies (Tab.~\ref{tab11}), the locations, distances and redshifts of the galaxies (Tab.~\ref{tab12}), apparent magnitudes of the galaxies (Tables~\ref{tab21} and~\ref{tab22}), luminosities of the galaxies (Tab.~\ref{tab31} and~\ref{tab32}), colours of the galaxies (Tables~\ref{tab41} and~\ref{tab42}), the structural properties of the galaxies (Tab.~\ref{tab51}) and the stellar populations of the galaxies (Tab.~\ref{tab52}). We have concentrated on collecting data on dwarf elliptical galaxies, for which some samples with detailed data have been published recently. Here, they have been united in a single catalogue. 

A potential problem with any catalogue that is collected from different sources is that they are inherently inhomogeneous. This catalogue is no different in that sense, even though we have performed some basic corrections in order to homogenise the data. However, on the other hand, this catalogue provides the largest collection of data on ETGs with intermediate luminosities and puts them into a context with the data on faint ETGs collected by \citet{McConnachie2012} and the data on massive ETGs provided by the ATLAS$^3D$ collaboration. Despite the limitations mentioned above, this catalogue should therefore be a useful resource for information on ETGs in the local Universe.

\section*{Acknowlegdements}
JD gratefully acknowledges funding through FONDECYT concurso postdoctorado grant~3140146. MF is funded by FONDECYT~1130521 and Basal-CATA. We also acknowledge that this research has made extensive use of several online tools and databases, namely the VizieR catalogue access tool, NASA's Astrophysics Data System, the NASA/IPAC Extragalactic Database (NED), the HyperLeda database (http://leda.univ-lyon1.fr), and SDSS-III. The original description of the VizieR service was published in A\&AS 143, 23. The NED is operated by the Jet Propulsion Laboratory, California Institute of Technology, under contract with the National Aeronautics and Space Administration. Funding for SDSS-III has been provided by the Alfred P. Sloan Foundation, the Participating Institutions, the National Science Foundation, and the U.S. Department of Energy Office of Science. The SDSS-III web site is http://www.sdss3.org/. SDSS-III is managed by the Astrophysical Research Consortium for the Participating Institutions of the SDSS-III Collaboration including the University of Arizona, the Brazilian Participation Group, Brookhaven National Laboratory, Carnegie Mellon University, University of Florida, the French Participation Group, the German Participation Group, Harvard University, the Instituto de Astrofisica de Canarias, the Michigan State/Notre Dame/JINA Participation Group, Johns Hopkins University, Lawrence Berkeley National Laboratory, Max Planck Institute for Astrophysics, Max Planck Institute for Extraterrestrial Physics, New Mexico State University, New York University, Ohio State University, Pennsylvania State University, University of Portsmouth, Princeton University, the Spanish Participation Group, University of Tokyo, University of Utah, Vanderbilt University, University of Virginia, University of Washington, and Yale University. We thank the referee, Dave Carter, for helpful comments. 

\bibliographystyle{mn2e}
\bibliography{ETGcatalog}

\label{lastpage}

\end{document}